\documentclass[twocolumn,showkeys,showpacs,preprintnumbers,prd,superscriptaddress,nofootinbib,aps,10pt]{revtex4-1}
\usepackage{graphicx,epsf,bm,amsmath,amsfonts,amssymb,epstopdf,natbib,verbatim,multirow,appendix,array,diagbox,xcolor,tikz,hyperref,booktabs,siunitx,float}
\usepackage[normalem]{ulem}
\hypersetup{colorlinks=true,urlcolor=blue,citecolor=blue,linkcolor=blue,menucolor=blue,anchorcolor=blue,filecolor=blue}

\newcolumntype{?}{!{\vrule width 3pt}}

\date{\today}

\makeatletter

\newcommand{\fmarki}{\ensuremath{\alpha}}
\newcommand{\fmarkii}{\ensuremath{\beta}}
\newcommand{\fmarkiii}{\ensuremath{\gamma}}
\newcommand{\fmarkiv}{\ensuremath{\delta}}
\newcommand{\fmarkv}{\ensuremath{\epsilon}}

\def\@fnsymbol#1{{\ifcase#1\or \fmarki\or \fmarkii\or \fmarkiii\or \fmarkiv\or \fmarkv\or \else\@ctrerr\fi}}
\makeatother
\definecolor{darkred}{rgb}{0.8, 0.0, 0.0}
\definecolor{darkgreen}{rgb}{0, 0.8, 0.0}
\definecolor{darkpink}{rgb}{0.8, 0.0, 0.4}

\definecolor{lime}{HTML}{A6CE39}
\DeclareRobustCommand{\orcidicon}{
\begin{tikzpicture}
\draw[lime, fill=lime] (0,0) 
circle [radius=0.16] 
node[white] {{\fontfamily{qag}\selectfont \tiny ID}};
\draw[white, fill=white] (-0.0625,0.095) 
circle [radius=0.007];
\end{tikzpicture}
\hspace{-3mm}
}

\foreach \x in {A, ..., Z}{\expandafter\xdef\csname orcid\x\endcsname{\noexpand\href{https://orcid.org/\csname orcidauthor\x\endcsname}{\noexpand\orcidicon}}}

\begin{document}

\title{Cosmological anatomy of interacting dark energy}

\author{\mbox{Miguel A. Sabogal\hspace{-1mm}\orcidA{}}}
\email{miguel.sabogalgarcia@unitn.it}
\affiliation{Department of Physics, University of Trento, Via Sommarive 14, 38123 Povo (TN), Italy}
\affiliation{Trento Institute for Fundamental Physics and Applications (TIFPA)-INFN, Via Sommarive 14, 38123 Povo (TN), Italy}

\author{\mbox{Sunny Vagnozzi\hspace{-1mm}\orcidB{}}}
\email{sunny.vagnozzi@unitn.it}
\affiliation{Department of Physics, University of Trento, Via Sommarive 14, 38123 Povo (TN), Italy}
\affiliation{Trento Institute for Fundamental Physics and Applications (TIFPA)-INFN, Via Sommarive 14, 38123 Povo (TN), Italy}

\author{\mbox{Eleonora Di Valentino\hspace{-1mm}\orcidC{}}}
\email{e.divalentino@sheffield.ac.uk}
\affiliation{School of Mathematical and Physical Sciences, University of Sheffield, Hounsfield Road, Sheffield S3 7RH, United Kingdom \looseness=-1}

\begin{abstract}
\noindent Interacting dark energy (IDE) models, featuring interactions between dark matter (DM) and dark energy (DE), have received significant renewed interest. However, little attention has been devoted to understanding their genuine cosmological signatures, with comparisons against $\Lambda$CDM often carried out at fixed cosmological parameters. Considering the widely studied model with energy exchange rate proportional to the DE density, we carry out a three-level sequence of comparisons, starting from the na\"{i}ve fixed-parameter comparison, then compensating for shifts in $\theta_s$ and $z_{\text{eq}}$, and then removing effects due to the modified background. We show that fixing $\theta_s$ and $z_{\text{eq}}$ leads to pre-recombination evolutions of the Weyl potential, photon monopole, and baryon velocity which are nearly identical to their $\Lambda$CDM counterparts, and hence virtually indistinguishable Cosmic Microwave Background (CMB) temperature power spectra. Comparing IDE to a non-interacting $w$CDM model with the same background, we find small scale-dependent signatures in CMB lensing. We find larger differences in the matter power spectrum, reflecting the different values of $\Omega_m$ resulting from the different partition of the dark sector into DM and DE, with only percent-level differences remaining once we compare $\Omega_m^2P_m(k)$, indicating the importance of high-fidelity determinations of $\Omega_m$. Our controlled comparison strategy can be a powerful tool for a wide range of models beyond $\Lambda$CDM.
\end{abstract}

\maketitle

\section{Introduction}
\label{sec:introduction}

Despite accounting for the overwhelming majority of the energy budget of the Universe, the microphysical nature of dark energy (DE) and dark matter (DM) remains unknown. Within the concordance $\Lambda$CDM model, DM and DE are assumed to interact only gravitationally, and their energy-momentum tensors are therefore separately (covariantly) conserved. However, if DM and DE arise from new (quantum) fields, non-gravitational interactions between them would generally be expected, unless forbidden or strongly suppressed by a fundamental symmetry~\cite{Carroll:1998zi}. Alongside attempts to address the coincidence problem, this was among the main theoretical motivations for introducing so-called interacting DE (IDE) models, which feature energy and momentum exchange between DM and DE (see e.g.\ Refs.~\cite{Wang:2016lxa,Wang:2024vmw} for reviews).~\footnote{Within these models, the coincidence problem is typically alleviated thanks to scaling or attractor solutions where the DM and DE densities remain comparable over an extended period. However, the amount of energy exchange which is required to address this problem is now understood to be excluded by cosmological observations, whereas further serious issues arise in relation to quantum corrections~\cite{DAmico:2016jbm,Marsh:2016ynw}.} While these models have a long history~\cite{Amendola:1999er,Mangano:2002gg,Farrar:2003uw}, observational interest in IDE has arguably boomed over the past decade (see e.g.\ Refs.~\cite{Skordis:2015yra,Li:2015vla,delCampo:2015vha,Casas:2015qpa,Nunes:2016dlj,Sebastiani:2016ras,Zhang:2017ize,Yang:2017zjs,Dutta:2017wfd,Kumar:2017bpv,Guo:2017deu,Dutta:2017fjw,Guo:2018gyo,Feng:2018yew,vonMarttens:2018iav,Li:2018ydj,Yang:2018qec,Yang:2019vni,Li:2019ajo,Carneiro:2019rly,Feng:2019jqa,Yang:2019uog,Vagnozzi:2019kvw,Pan:2020mst,Aljaf:2020eqh,Zhang:2021yof,Bonilla:2021dql,Xiao:2021nmk,Yang:2021oxc,Carrilho:2021hly,Ferlito:2022mok,Jin:2022tdf,Zhao:2022bpd,Hou:2022rvk,Li:2023gtu,Rodriguez-Benites:2023otm,Forconi:2023hsj,Halder:2024uao,Feng:2024lzh,Pooya:2024wsq,MV:2024vjv,Jimenez:2024lmm,Lima:2024wmy,vanderWesthuizen:2025mnw,vanderWesthuizen:2025vcb,vanderWesthuizen:2025rip,BeltranJimenez:2025yad,Silva:2025bnn,Wolney:2026zbc,Alvarez:2026kcr}): this is in no small part because of the emergence of a number of cosmological tensions, such as the Hubble tension~\cite{Verde:2019ivm,DiValentino:2020zio,DiValentino:2021izs,Perivolaropoulos:2021jda,Shah:2021onj,Abdalla:2022yfr,DiValentino:2022fjm,Hu:2023jqc,Vagnozzi:2023nrq,Verde:2023lmm,CosmoVerseNetwork:2025alb,Cai:2026swf,Schoneberg:2026eys,Schoneberg:2026vaf,Wang:2026jxd} and the $S_8$ tension~\cite{DiValentino:2018gcu,DiValentino:2020vvd,DES:2021bvc,DES:2021vln,DES:2021wwk,Nunes:2021ipq,Dalal:2023olq,Kilo-DegreeSurvey:2023gfr,Ghirardini:2024yni,ACT:2024okh,DES:2024oud,Wright:2025xka,DES:2026fyc,DES:2026mkc,Pantos:2026koc}, both of which, if not due to increasingly unlikely systematics~\cite{Efstathiou:2020wxn,Mortsell:2021nzg,Mortsell:2021tcx,Freedman:2021ahq,Kenworthy:2022jdh,Camarena:2022iae,Wojtak:2022bct,Riess:2023bfx,Giani:2023aor,Wojtak:2024mgg,Freedman:2024eph,Perivolaropoulos:2024yxv,Riess:2025chq,Hogas:2026urs,Sorrenti:2026tye}, may call for new physics beyond $\Lambda$CDM (see e.g.\ Refs.~\cite{Anchordoqui:2015lqa,DiValentino:2016hlg,Karwal:2016vyq,Vagnozzi:2017ovm,Mortsell:2018mfj,Vagnozzi:2018jhn,Poulin:2018dzj,RoyChoudhury:2018gay,Guo:2018ans,Poulin:2018cxd,Kreisch:2019yzn,Agrawal:2019lmo,Lin:2019qug,Yang:2019nhz,Vagnozzi:2019ezj,Visinelli:2019qqu,Smith:2019ihp,Niedermann:2019olb,Berghaus:2019cls,Sakstein:2019fmf,Hart:2019dxi,Nojiri:2019fft,Ye:2020btb,Krishnan:2020obg,Zumalacarregui:2020cjh,Ballesteros:2020sik,Alestas:2020mvb,Jedamzik:2020krr,Braglia:2020iik,Ballardini:2020iws,Gogoi:2020qif,Braglia:2020bym,Gonzalez:2020fdy,Sekiguchi:2020teg,Ye:2020oix,Niedermann:2020qbw,Murgia:2020ryi,Smith:2020rxx,CarrilloGonzalez:2020oac,Braglia:2020auw,Adi:2020qqf,Oikonomou:2020qah,Oikonomou:2020oex,RoyChoudhury:2020dmd,Brinckmann:2020bcn,Marra:2021fvf,SolaPeracaula:2021gxi,Dainotti:2021pqg,Teng:2021cvy,Krishnan:2021dyb,Vagnozzi:2021tjv,Vagnozzi:2021gjh,Jiang:2021bab,Gomez-Valent:2021cbe,Hart:2021kad,Ye:2021iwa,Cyr-Racine:2021oal,Akarsu:2021fol,Niedermann:2021ijp,Niedermann:2021vgd,Wang:2022jpo,Nojiri:2022ski,Oikonomou:2022yle,Schoneberg:2022grr,Reeves:2022aoi,RoyChoudhury:2022rva,Moshafi:2022mva,Rezazadeh:2022lsf,Escudero:2022rbq,Banerjee:2022ynv,deSa:2022hsh,Akarsu:2022typ,Lee:2022gzh,Khodadi:2023ezj,Ben-Dayan:2023rgt,Poulin:2023lkg,Cruz:2023lmn,Gomez-Valent:2023hov,Odintsov:2023cli,Ruchika:2023ugh,Adil:2023exv,Frion:2023xwq,Akarsu:2023mfb,Sharma:2023kzr,Gomez-Valent:2023uof,Efstathiou:2023fbn,Pedreira:2023qqt,Khalife:2023qbu,Akarsu:2024qiq,Erdem:2024vsr,Giare:2024akf,Lynch:2024hzh,Yadav:2024duq,Toda:2024ncp,Nozari:2024wir,Escamilla:2024xmz,Chatrchyan:2024xjj,RoyChoudhury:2024wri,Jiang:2024nha,Mirpoorian:2024fka,Gomez-Valent:2024ejh,Jiang:2025ylr,Stahl:2025czl,Akarsu:2025gwi,Jiang:2025hco,Lee:2025yah,Poulin:2025nfb,Toda:2025kcq,Wang:2025dzn,Erdem:2025xtr,Efstratiou:2025iqi,Chen:2025nfy,Kumar:2025obb,Hogas:2025mii,Montani:2025nmz,Pantos:2026cxv,Bouhmadi-Lopez:2026dte,Wang:2026rkx,Gonzalez-Fuentes:2026rgu,Mukhopadhyay:2026fyk,Pantos:2026rpe,Bella:2026zuk,Carloni:2026yut,Pedrotti:2026dwj,Wang:2026kor,Jusufi:2026rfd,Dhyani:2026trw,Giare:2026tyk,Li:2026hwq,Lee:2026hlo,Akarsu:2026lva,Jia:2026vdt,Du:2026qtq,Li:2026asg,Hashim:2026yoy,Sabogal:2026qvy,Bouhmadi-Lopez:2026vyc,Tiwari:2026pzk,Ajith:2026qqq,KumarSharma:2026ptx,DOnofrio:2026dyq,SolaPeracaula:2026lyk,Tsilioukas:2026gvy,Chaudhary:2026hwq,Chaudhary:2026ybu,Zhou:2026jmn,Jia:2026ktx,Chaudhary:2026wuq,Sengupta:2026gyg,Banihashemi:2026twb}). Despite well-known no-go arguments against purely late-time solutions to the Hubble tension~\cite{Bernal:2016gxb,Lemos:2018smw,Aylor:2018drw,Schoneberg:2019wmt,Knox:2019rjx,Arendse:2019hev,Efstathiou:2021ocp,Cai:2021weh,Keeley:2022ojz,Jiang:2024xnu,Zhou:2025kws,Pedrotti:2025ccw,Bansal:2026axl,Zhou:2026iar,Sabogal:2026ipu}, IDE models are still relevant in this context because the exchange of energy and momentum between DM and DE modifies both the background expansion history and the growth of structure, affecting the observables relevant to both tensions, which IDE is able to partially alleviate (see for instance Refs.~\cite{Murgia:2016ccp,Pourtsidou:2016ico,Kumar:2016zpg,Guo:2017hea,Yang:2017ccc,An:2017crg,Barros:2018efl,Yang:2018euj,Yang:2018uae,vonMarttens:2018bvz,Martinelli:2019dau,Kumar:2019wfs,Pan:2019jqh,Yang:2019uzo,DiValentino:2019ffd,Benetti:2019lxu,DiValentino:2019jae,Pan:2020zza,Lucca:2020zjb,Hogg:2020rdp,Gomez-Valent:2020mqn,DiValentino:2020evt,DiValentino:2020leo,DiValentino:2020vnx,DiValentino:2020kpf,BeltranJimenez:2020qdu,Gao:2021xnk,Wang:2021kxc,Benetti:2021div,Kumar:2021eev,Figueruelo:2021elm,Jimenez:2021ybe,Karwal:2021vpk,Linton:2021cgd,Nunes:2021zzi,Anchordoqui:2021gji,Hogg:2021yiz,Guo:2021rrz,Gariazzo:2021qtg,Yao:2020hkw,Yao:2022kub,Gomez-Valent:2022bku,Bernui:2023byc,Zhai:2023yny,Hoerning:2023hks,Liu:2023rvo,Giare:2024ytc,Pookkillath:2024roo,Yang:2025vnm,Ambelu:2026mkd} for various studies in this direction). The first data releases from the Dark Energy Spectroscopic Instrument (DESI) have further stimulated interest in IDE models over the past two years, with several studies reporting tentative indications of DM-DE interactions~\cite{Giare:2024smz,Li:2024qso,Silva:2025hxw,Pan:2025qwy,Li:2026xaz} (see also Refs.~\cite{Chakraborty:2024xas,Montani:2024pou,Sabogal:2024yha,Ghedini:2024mdu,Aboubrahim:2024cyk,Li:2025owk,Sabogal:2025mkp,Nagpal:2025rnu,Feng:2025mlo,Chakraborty:2025syu,Zhai:2025hfi,Shah:2025ayl,You:2025uon,Yashiki:2025loj,Li:2025ula,Guedezounme:2025wav,Arora:2025ecj,Petri:2025swg,Yang:2025uyv,Hussain:2025uye,Samanta:2025oqz,Li:2025muv,Zhang:2025dwu,Gonzalez-Espinoza:2025vrc,Wu:2025vrl,Paliathanasis:2026ymi,Figueruelo:2026eis,Escobal:2026zxb,Aboubrahim:2026tks,Escobal:2026lnp,Dai:2026pvx,Kashyap:2026ivg,Wang:2026wrk,Antusch:2026ldp,Paliathanasis:2026acn,Neumann:2026qpq,Artola:2026tgs,Wang:2026vqw,Abdalla:2026sis,Huang:2026sip,Li:2026ldf,Zhai:2026uwr,Wang:2026gvg,Kandel:2026uhe,Yang:2026vkc,Pompeu:2026zad,Wang:2026esu,Zhang:2026dtd,Sahlu:2026cib,Yang:2026qmz}). Although far from conclusive, these recent developments make it especially timely to understand \textit{precisely} what the cosmological signatures of IDE are. As we shall see, answering this (only apparently) simple question is far from straightforward, and to do so is the main goal of our work.

Despite this vast literature, observational studies of IDE have almost exclusively focused on parameter inference, asking, for instance, whether cosmological data indicate a non-zero coupling and whether this can alleviate cosmological tensions, or, conversely, how tightly current and future data can constrain the interaction strength, typically denoted by $\xi$. Significantly less attention has been paid to a \textit{careful} study of the \textit{physical} origin of the observable signatures being sought in the data. To clarify our emphasis on the adjectives ``careful'' and ``physical'', we note that, when the effects of IDE on Cosmic Microwave Background (CMB) anisotropies and Large-Scale Structure (LSS) clustering are shown, this is usually done by varying $\xi$ while keeping all other cosmological parameters fixed.~\footnote{Concrete examples of these types of comparison plots are, for instance, Fig.~5 of Ref.~\cite{Yang:2018euj} for the CMB temperature power spectrum, and Fig.~1 of Ref.~\cite{Nunes:2022bhn} for the galaxy power spectrum multipoles. To avoid singling out works by others while illustrating a widespread practice, we have deliberately drawn both examples from works coauthored by some of us (E.D.V.\ and S.V.). We stress that our point concerns solely the interpretation of these illustrative plots, and not the validity of the parameter inference analyses performed in these works.} As a first test, this is useful for gaining a feeling for the overall response of the CMB and matter power spectra to $\xi$. However, such \textit{fixed-parameter} analyses cannot discriminate between features which are \textit{intrinsic} to the DM-DE interaction and those which can (and, in general, will) be absorbed by shifts in other parameters. These shifts are required to keep a number of accurately measured scales fixed, as we shall discuss shortly. Partial steps in this direction have been taken, most notably in Refs.~\cite{Lucca:2021dxo,Lucca:2021eqy,Petri:2025swg}. However, to the best of our knowledge, there is no comprehensive analysis of IDE which has systematically separated effects arising from shifts in cosmological scales, modified background evolution, and genuine interaction signatures across CMB and LSS observables. For the reasons highlighted previously, closing this gap is a timely endeavor and is the main motivation for our work.

The main reason why fixed-parameter comparisons can be misleading is that varying $\xi$ affects not only the evolution of perturbations, but also the background expansion rate. As a result, when holding all other cosmological parameters fixed, varying $\xi$ shifts a number of accurately measured quantities, for instance the acoustic angular scale $\theta_s$ and the redshift of matter-radiation equality $z_{\text{eq}}$. These shifts lead to prominent but disproportionately large effects on the CMB and matter power spectra, which would be excluded at extremely high significance by cosmological observations, even for values of $\xi$ which are actually allowed by parameter inference analyses. In fact, in an actual parameter inference analysis, these large shifts are to a significant extent compensated by correlated shifts in other cosmological parameters, which restore, among others, $\theta_s$ and $z_{\text{eq}}$ to their original values: the resulting effects cannot therefore be considered distinctive signatures of IDE. We believe that a controlled comparison, at least in the context of IDE, must proceed in two stages, beyond from the (uninformative) fixed-parameter one. In a first \textit{scale-matched} comparison, one should compensate for shifts in the relevant well-measured scales by adjusting the remaining cosmological parameters while $\xi$ is varied. The differences which survive this stage are physically meaningful and, most importantly, observationally relevant signatures of IDE. However, they still contain the combined effects of \textit{a)} the modified background evolution and \textit{b)} perturbation-level signatures of the DM-DE interaction. The former can be, and indeed is, constrained by background probes such as Baryon Acoustic Oscillation (BAO) and Type Ia Supernovae (SNeIa) measurements, but cannot be considered a distinctive signature of dark sector interactions. For this reason, it is desirable to disentangle these two contributions, and this in turn requires comparison with a non-interacting cosmology with the same background history as the IDE model under consideration: it is only the differences surviving this second \textit{expansion-matched} stage which can be cleanly identified as genuine signatures of IDE, and it is these differences we are mostly after. We stress that these nuances, and more generally the importance of distinguishing fixed-parameter effects from true underlying physical effects, are well known and appreciated in other contexts, for instance neutrino cosmology (see, for example, the beautiful discussions in Chapters~5 and~6 of Ref.~\cite{Lesgourgues:2013sjj}): however, they have received much less attention in the context of IDE.

In this work, focusing on one of the simplest and most widely tested phenomenological IDE models, with energy exchange rate proportional to the expansion rate and the DE density (see e.g.\ Ref.~\cite{DiValentino:2019ffd}), we carry out the controlled comparisons described above to dissect (hence the use of the word ``anatomy'' in the title) the physical origin of IDE's cosmological signatures. We start by illustrating the conventional but potentially misleading fixed-parameter comparison. We then move on to a scale-matched comparison, where interaction-induced shifts in key cosmological scales are compensated by shifts in other parameters. Finally, we carry out an expansion-matched comparison, where the scale-matched IDE cosmologies are compared to non-interacting $w$CDM models with the same background expansion history. This sequence of comparisons allows us to progressively separate signatures associated with shifts in accurately measured cosmological scales from those due to differences in the background evolution, and finally to isolate genuine IDE signatures which survive both comparisons. We carry out this analysis focusing on both CMB and LSS observables, carefully dissecting the individual physical contributions to the CMB power spectrum. As a brief appetizer for our main results, we find that, once the hierarchy of comparisons is completed, the primary CMB spectra of the models being compared are largely indistinguishable. Small but non-vanishing differences remain in the CMB lensing power spectrum, while much larger differences survive in the matter power spectrum, although these are mostly accounted for by the different values of $\Omega_m$ in the cosmologies being compared: when instead comparing $\Omega_m^2P_m(k)$, only a percent-level, mildly scale-dependent signature remains. Overall, our results show that IDE's cleanest signatures are those related to the modified background evolution, whereas the perturbation-level signatures which survive our controlled comparisons are substantially smaller, at least for the model studied here.

The rest of this paper is organized as follows. In Sec.~\ref{sec:ide}, we begin by briefly reviewing the IDE model considered here, introducing all the relevant equations. Sec.~\ref{sec:models} is instead devoted to discussing in detail the three levels of controlled comparisons we carry out. After briefly illustrating the conventional but otherwise misleading fixed-parameter comparison, in Sec.~\ref{sec:scalematched} we perform the scale-matched comparison, whereas the expansion-matched comparison is carried out in Sec.~\ref{sec:genuineeffects}. A critical discussion of our results and their broader implications is then provided in Sec.~\ref{sec:discussion}. Finally, we conclude in Sec.~\ref{sec:conclusions}.

\section{Interacting dark energy}
\label{sec:ide}

We now briefly introduce the IDE model which will be dissected throughout this work. In what follows, we work within a spatially flat Friedmann-Lema\^{\i}tre-Robertson-Walker background. Rather than starting from first principles within a specific particle model, we introduce a phenomenological parametrization of the DM-DE interaction at the level of the conservation equations for the respective stress-energy tensors, $T^{\mu\nu}_c$ and $T^{\mu\nu}_x$. Hereafter, the subscripts $c$ and $x$ will always refer to DM and DE respectively. As is widely done in the literature, we assume that, unlike within the $\Lambda$CDM model where the two stress-energy tensors are separately (covariantly) conserved, here only their sum is. Specifically, we assume that the covariant derivatives of the stress-energy tensors evolve as follows:
\begin{equation}
\nabla_{\nu}T^{\mu\nu}_c=\frac{Qu^{\mu}_c}{a}\,, \quad \nabla_{\nu}T^{\mu\nu}_x=-\frac{Qu^{\mu}_c}{a}\,.
\label{eq:nablatmunu}
\end{equation}
where $a$ is the scale factor, $Q$ is the interaction rate (energy exchanged per unit volume per unit time), and $u^{\mu}_c$ is the DM four-velocity. Within the above parametrization, the energy-momentum transfer four-vector is parallel to $u^{\mu}_c$: as a consequence, there is no momentum transfer in the DM rest frame, and the DM Euler equation remains unmodified. With regard to the interaction rate $Q$, a phenomenological choice which is widely adopted in the literature is the following:
\begin{equation}
Q=\xi{\cal H}\rho_x\,,
\label{eq:q}
\end{equation}
where ${\cal H}$ is the conformal Hubble rate, $\rho_x$ is the DE energy density, and the dimensionless parameter $\xi$ is the interaction strength. With the parametrization of Eqs.~(\ref{eq:nablatmunu},\ref{eq:q}), a positive (negative) value of $\xi$ corresponds to energy transfer from DE to DM (from DM to DE). One may wonder whether the explicit dependence of $Q$ on ${\cal H}$ is physically justified, as the microscopic interaction between DM and DE is ultimately expected to be local, whereas the expansion rate is a global quantity. This apparent tension can be understood from the first law of thermodynamics~\cite{Nunes:2022bhn}:~\footnote{E.D.V.\ and S.V.\ thank the anonymous reviewer of one of their previous papers for drawing their attention to this important issue, whereas S.V.\ thanks Marco Bruni for providing this illuminating explanation.} the conservation equations locally relate changes in energy density to changes in the physical volume as the Universe expands. Seen this way, the appearance of ${\cal H}$ simply reflects the rate at which this volume changes, and not a direct sensitivity of local interactions to the global cosmological background. In fact, the same conservation equations can be rewritten using the scale factor as time variable, making the explicit dependence on ${\cal H}$ disappear. Furthermore, we stress that similar interaction rates arise from well-motivated IDE field theory constructions~\cite{Pan:2020zza}.

Assuming for simplicity that the DE component is characterized by a constant equation of state (EoS) $w_x \neq -1$, Eqs.~(\ref{eq:nablatmunu},\ref{eq:q}) lead to the following continuity equations for the DM and DE energy densities, $\rho_c$ and $\rho_x$:
\begin{align}
\dot{\rho}_c+3{\cal H}\rho_c&= \xi{\cal H}\rho_x\,,
\label{eq:continuityc}\\
\dot{\rho}_x+3{\cal H}(1+w_x)\rho_x&= -\xi{\cal H}\rho_x\,.
\label{eq:continuityx}
\end{align}
Since $w_x$ is constant, the above equations can be easily integrated analytically to obtain the evolution of $\rho_c$ and $\rho_x$ as a function of the scale factor:
\begin{align}
\rho_c&= \frac{\rho_{c,0}}{a^3}+\frac{\rho_{x,0}}{a^3} \left [ \frac{\xi}{3w_x+\xi} \left ( 1-a^{-3w_x-\xi} \right ) \right ] \,,
\label{eq:solutionrhoc}\\
\rho_x&= \frac{\rho_{x,0}}{a^{3(1+w_x)+\xi}}\,,
\label{eq:solutionrhox}
\end{align}
where $\rho_{c,0}$ and $\rho_{x,0}$ are the present-day DM and DE energy densities. From the latter expression, we see that the background evolution of the DE energy density is equivalent to that of a non-interacting component whose effective EoS is the following:
\begin{equation}
w_{x,\text{eff}}=w_x+\frac{\xi}{3}\,.
\end{equation}

To treat linear perturbations, we work in synchronous gauge, where the equations for the DM (DE) density contrast and velocity divergence, $\delta_c$ and $\theta_c$ ($\delta_x$ and $\theta_x$), take the following form~\cite{Gavela:2010tm}:
\begin{align}
\dot{\delta}_c&= -\theta_c-\frac{1}{2}\dot{h}+\xi{\cal H}\frac{\rho_x}{\rho_c} \left ( \delta_x-\delta_c \right ) +\xi\frac{\rho_x}{\rho_c} \left ( \frac{kv_T}{3}+\frac{\dot{h}}{6} \right ) \,,
\label{eq:deltac}\\
\dot{\theta}_c&= -{\cal H}\theta_c\,,
\label{eq:thetac}\\
\dot{\delta}_x&= -(1+w_x) \left ( \theta_x+\frac{\dot{h}}{2} \right )-\xi \left ( \frac{kv_T}{3}+\frac{\dot{h}}{6} \right ) \nonumber\\
&\quad-3{\cal H}(1-w_x) \left [ \delta_x+\frac{{\cal H}\theta_x}{k^2} \left ( 3+3w_x+\xi \right ) \right ]\,,
\label{eq:deltax}\\
\dot{\theta}_x&= 2{\cal H}\theta_x+\frac{k^2}{1+w_x}\delta_x \nonumber\\
&\quad+2{\cal H}\frac{\xi}{1+w_x}\theta_x-\xi{\cal H}\frac{\theta_c}{1+w_x}\,,
\label{eq:thetax}
\end{align}
where dots denote derivatives with respect to conformal time, $h$ is the trace of the synchronous-gauge scalar metric perturbation, and $v_T$ is the center-of-mass velocity of the total fluid. Moreover, as commonly done, we set the DE rest-frame sound speed squared to $c_{s,x}^2=1$: this provides strong pressure support against gravitational collapse and therefore suppresses DE clustering on sub-horizon scales. We note that the equation for $\theta_c$ is unmodified as a consequence of our assumption that the energy-momentum transfer four-vector is parallel to the DM four-velocity. Finally, we set adiabatic initial conditions for $\delta_x$ and $\theta_x$ following Ref.~\cite{Gavela:2010tm}. As is well known, gravitational and non-adiabatic instabilities in this model can be avoided provided $w_x \neq -1$ and the signs of $(1+w_x)$ and $\xi$ are opposite. To derive predictions for cosmological observables, the above modifications are implemented in the \texttt{CLASS} Boltzmann solver~\cite{Blas:2011rf}.~\footnote{Our custom version of \texttt{CLASS} is publicly available at \href{https://github.com/msabogal/CLASS\_IDE}{https://github.com/msabogal/CLASS\_IDE}.}

As we will discuss in more detail in Sec.~\ref{sec:models}, in this work we consider two different classes of models. In the first, we set $w_x=-0.999$ and therefore require $\xi<0$, with the goal of studying a model which is as close as possible to an interacting vacuum scenario (for which strictly $w_x=-1$). Taking $1+w_x$ to be small but non-zero allows us to approach this limit while avoiding the singular behavior of the fluid perturbations for $w_x=-1$, while strongly suppressing the impact of DE perturbations on the Einstein-Boltzmann system. Such a phenomenological prescription, explicitly demonstrated to work by one of us in Ref.~\cite{DiValentino:2020leo}, has found widespread use in the recent literature on IDE (see for instance Refs.~\cite{DiValentino:2019ffd,Lucca:2020zjb,Xiao:2021nmk,Hoerning:2023hks,Rodriguez-Benites:2023otm}). In the second class of models, we instead set $w_x=-0.8$, while still requiring $\xi<0$. In both cases, we focus on negative values of $\xi$, since these are favored by current data~\cite{Giare:2024smz,Li:2024qso,Silva:2025hxw,Pan:2025qwy,Li:2026xaz}: with our sign convention, these lead to energy transfer from DM to DE. The specific model considered here, while phenomenological in nature, has been extensively studied and constrained against cosmological data over the past decade. We stress that the phenomenological approach adopted here should be distinguished from first-principles approaches, where the energy-momentum transfer is derived from an underlying Lagrangian rather than specified directly at the level of the conservation equations~\cite{Pan:2020zza,Aboubrahim:2024cyk,Zhang:2025dwu,Li:2026xaz,Abdalla:2026sis}.

To characterize the impact of the DM-DE interaction on the evolution of matter perturbations, particularly on linear scales, we extend the approximate non-relativistic sub-horizon treatment adopted in Ref.~\cite{Silva:2024ift} for CDM to the total matter sector, defining $\rho_{\rm m}\delta_{\rm m}=\rho_{\rm b}\delta_{\rm b}+\rho_{\rm c}\delta_{\rm c}$. This extension is possible because DE interacts only with DM, while the baryonic sector remains minimally coupled, as in $\Lambda$CDM. For this purpose, we temporarily work in Newtonian gauge,~\footnote{On sub-horizon scales, the differences between density contrasts in the two gauges are negligible.} where the Poisson equation is given by the following:
\begin{equation}
-k^2\psi=4\pi G a^2\rho_m\delta_m\,,
\label{eq:poisson}
\end{equation}
and $\psi$ corresponds to the Newtonian gauge gravitational potential. If we define the total matter growth factor as $D(a)\equiv\delta_{\rm m}(a)/\delta_{\rm m}(1)$, we can combine the matter continuity and Euler equations with Eq.~(\ref{eq:poisson}) to obtain the following growth equation:~\footnote{Extending the growth equation from DM alone to the total matter sector would introduce a residual term $R$ on the right-hand side, $R=\xi\mathcal{H}r_{xm}f_bC(\tau_{\text{ini}})/a$, where $f_b\equiv\rho_b/\rho_m$ and $C(\tau_{\rm ini})=a(\tau_{\text{ini}})[\dot{\Delta}(\tau_{\text{ini}})+\xi\mathcal{H}(\tau_{\text{ini}})r_{xc}(\tau_{\rm ini})\delta_c(\tau_{\text{ini}})]$, with $\Delta\equiv\delta_c-\delta_b$ and $r_{xc} \equiv \rho_x/\rho_c$. This residual term vanishes if $C(\tau_{\text{ini}}) \simeq 0$. This condition is satisfied to very good approximation for adiabatic initial conditions, for which $\Delta(\tau_{\text{ini}})\simeq\dot{\Delta}(\tau_{\text{ini}})\simeq0$, taking into account the fact that the interaction contribution $\xi r_{xc}$ is negligible at sufficiently early times, when $\rho_x\ll\rho_c$. Under this standard and well-justified assumption, $R \simeq 0$, and Eq.~(\ref{eq:growth}) remains valid to very good approximation for the total matter sector.}
\begin{equation}
\ddot{D}+\mathcal{H} \left [ 1+r_{xm}\xi \right ] \dot{D} -D \left [ 4\pi G a^2\rho_m+S(\tau,\xi) \right ] =0\,,
\label{eq:growth}
\end{equation}
where $r_xm\equiv\rho_x/\rho_m$, and the function $S(\tau,\xi)$ is given by the following:
\begin{equation}
S(\tau,\xi)=r_{xm}\xi \left [ \mathcal{H}^2 \left ( 3w_{x,\text{eff}}+r_{xm}\xi-1 \right ) -\dot{\mathcal{H}} \right ] \,.
\label{eq:stauxi}
\end{equation}
We therefore see that the DM-DE interaction modifies both the friction experienced by total matter perturbations (the term proportional to $\dot{D}$) and the source driving their growth (through the $S(\tau,\xi)$ term). The standard $\Lambda$CDM total matter growth equation is recovered in the limit $\xi\to0$ and $w_{\rm x}\to-1$. We stress that these approximate expressions are provided only to aid physical intuition and are not used in any subsequent calculation. All our numerical predictions are obtained by solving the complete Einstein-Boltzmann system, including the baryonic and DE components and their perturbations.

\section{Models and comparison strategy}
\label{sec:models}

\begin{table*}[!htb]
\centering
\resizebox{0.8\textwidth}{!}{
\begin{tabular}{|c?c||c|c|c||c|c|}
\hline
\multirow{2}{*}{\textbf{Parameter}} & \multirow{2}{*}{$\Lambda$CDM} & \multicolumn{3}{c||}{\textbf{Class $A$}} & \multicolumn{2}{c|}{\textbf{Class $B$}} \\
\cline{3-7}
& & IDE$_{\text{FP}}^A$ & IDE$_{\text{SM}}^A$ & $w$CDM$_{\text{EM}}^A$ & IDE$_{\text{SM}}^B$ & $w$CDM$_{\text{EM}}^B$ \\
\hline\hline
$100\theta_s$ & $1.044235$ & $1.062346$ & Ref. & Ref. & Ref. & Ref. \\
$\ln(10^{10}A_s)$ & $3.044$ & Ref. & Ref. & Ref. & Ref. & Ref. \\
$\tau_{\text{reio}}$ & $0.0546$ & Ref. & Ref. & Ref. & Ref. & Ref. \\
$n_s$ & $0.966$ & Ref. & Ref. & Ref. & Ref. & Ref. \\
$\omega_b$ & $0.022698$ & Ref. & Ref. & Ref. & Ref. & Ref. \\
$\omega_c$ & $0.121798$ & Ref. & $0.085321$ & $0.121799$ & $0.086594$ & $0.121801$ \\
$\xi$ & -- & $-0.3$ & $-0.3$ & -- & $-0.3$ & -- \\
$w_x$ & $-1$ & $-0.999$ & $-0.999$ & $-1.099$ & $-0.8$ & $-0.9$ \\
\hline
$\Omega_m$ & $0.309$ & Ref. & $0.212$ & $0.283$ & $0.256$ & $0.338$ \\
$z_{\text{eq}}$ & $3453.7$ & $4139.6$ & Ref. & Ref. & Ref. & Ref. \\
$H_0\,[{\text{km}}/{\text{s}}/{\text{Mpc}}]$ & $68.28$ & Ref. & $71.35$ & $71.35$ & $65.30$ & $65.30$ \\
\hline
\end{tabular}}
\caption{Parameters defining the six reference cosmologies used in our main sequence of controlled comparisons. The labels FP, SM, and EM denote fixed-parameter, scale-matched, and expansion-matched cosmologies respectively, while classes $A$ and $B$ correspond to values of the DE EoS given by $w_x=-0.999$ and $-0.8$ respectively. ``Ref.'' indicates that the quantity is fixed to its value in the reference $\Lambda$CDM cosmology.}
\label{tab:referencecosmologies}
\end{table*}

We now define the different reference cosmologies and the controlled comparison strategy we adopt to extract the physical effects of IDE on the cosmological observables of interest. The cosmological parameters defining these six cosmologies are summarized in Tab.~\ref{tab:referencecosmologies}. Throughout our work, the values of the physical baryon density $\omega_b$, the optical depth to reionization $\tau$, and the amplitude and tilt of the primordial scalar power spectrum, $A_s$ and $n_s$, are always fixed to those of our reference $\Lambda$CDM cosmology, itself close to the \textit{Planck} 2018 best-fit cosmology~\cite{Planck:2018vyg}. In this way, we ensure that differences between the cosmologies being compared do not arise from changes in the primordial power spectrum, reionization history, or baryonic content. As we will explain in more detail later, we further separate the IDE cosmologies and their non-interacting $w$CDM counterparts into two classes, named $A$ and $B$: these are distinguished by the values of the DE EoS $w_x$, and consequently by their effective EoS $w_{x,\mathrm{eff}}$.

We proceed through the following sequence of controlled comparisons:
\begin{enumerate}
\item The first question we ask is what happens if we simply switch on the DM-DE interaction while keeping all the other input cosmological parameters, including $H_0$, fixed to their reference $\Lambda$CDM values.~\footnote{Whether $H_0$ or $\theta_s$ is treated as an input parameter is merely a choice of parameter basis. We opt for holding $H_0$ fixed since most of the original fixed-parameter comparisons were carried out using \texttt{CAMB}, which requires $H_0$ as a user-supplied input. On the other hand, \texttt{CLASS} accepts either $H_0$ or $\theta_s$ as input.} This question is addressed by comparing the $\Lambda$CDM model against the IDE$_{\text{FP}}^A$ model, where ``FP'' stands for fixed-parameter, and $A$ labels the model class (as discussed in more detail later).
\item The second question we ask is which signatures remain once we preserve the acoustic angular scale $\theta_s$ and the redshift of matter-radiation equality $z_{\text{eq}}$, which we achieve through shifts in other cosmological parameters. The importance of holding these two scales fixed will be explained shortly. This question is addressed by comparing the $\Lambda$CDM model against the IDE$_{\text{SM}}^A$ model, where ``SM'' stands for scale-matched.
\item The third question we ask is which signatures remain when the scale-matched IDE model is compared against a non-interacting $w$CDM model with exactly the same background expansion history. This question is addressed by comparing the IDE$_{\text{SM}}^A$ model against the $w$CDM$_{\text{EM}}^A$ model, where ``EM'' stands for expansion-matched. Since residual differences between these models cannot be attributed to differences in their background expansion histories, they can be identified as genuine interaction signatures.
\item Finally, we repeat both the scale-matched and expansion-matched comparisons for class $B$, comparing first the $\Lambda$CDM model against the IDE$_{\text{SM}}^B$ model, and then the IDE$_{\text{SM}}^B$ model against the $w$CDM$_{\text{EM}}^B$ model.
\end{enumerate}
The above sequence of comparisons takes us from the ``raw'' but otherwise misleading effects of na\"{i}vely switching on the interaction, to the effects which remain after removing shifts in the two well-constrained physical scales, and finally to the genuine effects of DM-DE interactions which cannot be attributed to differences in the background expansion history.

Whenever a parameter is indicated by ``Ref.'' in Tab.~\ref{tab:referencecosmologies}, its value is fixed to the reference $\Lambda$CDM value. Starting from this reference cosmology, the IDE$_{\text{FP}}^A$ model is defined by a DM-DE interaction strength $\xi=-0.3$. This relatively large value of $\xi$ is chosen deliberately to amplify IDE's signatures and make their physical origin easier to visualize and identify, while remaining broadly compatible with current cosmological constraints. On the other hand, $H_0$, the physical DM density $\omega_c$, and all other parameters are fixed to their reference values. As for $H_0$, we provide it as an input parameter to \texttt{CLASS}, which therefore determines $\theta_s$ from the resulting background evolution. Finally, as stressed earlier, within model class $A$ we fix $w_x=-0.999$, which allows us to explore the interacting vacuum limit while avoiding the singular behavior of fluid perturbations for $w_x=-1$, a phenomenological prescription which has been explicitly demonstrated to work in Ref.~\cite{DiValentino:2020leo}. As is clear from Tab.~\ref{tab:referencecosmologies}, within the IDE$_{\text{FP}}^A$ model both $\theta_s$ and $z_{\text{eq}}$ shift significantly compared to their reference values.

The large differences we will observe between the $\Lambda$CDM and IDE$_{\text{FP}}^A$ models within this first \textit{fixed-parameter} comparison, to be discussed at the start of Sec.~\ref{sec:scalematched}, are not truly representative of IDE. The reason is precisely the large shifts in $\theta_s$ and $z_{\text{eq}}$ discussed above. Since CMB observations are extremely sensitive to these characteristic scales, it is somewhat unwise to vary $\xi$ without compensating for the shifts in these scales. Specifically, $\theta_s$ is one of the very few cosmological parameters which is measured with almost no model dependence, as it is fixed to exquisite precision by the position and spacing of the CMB acoustic peaks.\footnote{This statement assumes a description of pre-recombination physics which is very close to the standard one. In more general scenarios, the inferred value of $\theta_s$ can shift because of degeneracies with parameters which modify the acoustic peak structure. An important example is provided by modifications to neutrino free-streaming, which alter the neutrino-induced phase shift of the CMB acoustic peaks. The resulting change in the peak positions can be compensated by shifts in $\theta_s$ and other parameters, allowing values as large as $100\theta_s \simeq 1.046$, while providing a good fit to CMB data~\cite{Kreisch:2019yzn}.} On the other hand, $z_{\text{eq}}$ controls the amount of expansion between matter-radiation equality and decoupling. During this period, gravitational potentials decay, sourcing temperature anisotropies through the early integrated Sachs-Wolfe effect and mainly boosting the first acoustic peak. The exquisitely measured height of the first acoustic peak is therefore very sensitive to $z_{\text{eq}}$. Aside from this, $z_{\text{eq}}$ is in principle constrained by the turnover in the matter power spectrum. For these reasons, it is a good idea to study the effects of IDE by turning on $\xi$ while simultaneously compensating for the shifts in $\theta_s$ and $z_{\text{eq}}$ through shifts in other parameters: this is the typical procedure followed, for instance, when studying massive neutrino signatures in the CMB~\cite{Lesgourgues:2013sjj,Vagnozzi:2019utt}. In what follows, we therefore define the \textit{scale-matched} IDE$_{\text{SM}}^A$ model, which has the same values $\xi=-0.3$ and $w_x=-0.999$, but is constructed precisely so as to preserve the reference values of $\theta_s$ and $z_{\text{eq}}$. In practice, we achieve this by providing $\theta_s$, as opposed to $H_0$, as an input to \texttt{CLASS}, while adjusting $\omega_c$ in such a way as to keep $z_{\text{eq}}$ fixed. Of course, when doing so, $H_0$ is no longer independently fixed but is determined by \texttt{CLASS}, and will therefore shift relative to its reference value. Comparing the $\Lambda$CDM and IDE$_{\text{SM}}^A$ models within this scale-matched comparison (to be discussed in Sec.~\ref{sec:scalematched}) allows us to extract the interaction signatures which survive once shifts in important characteristic cosmological scales are appropriately reabsorbed by shifts in other cosmological parameters.

\begin{figure*}[!htpb]
\centering
\includegraphics[width=0.99\linewidth]{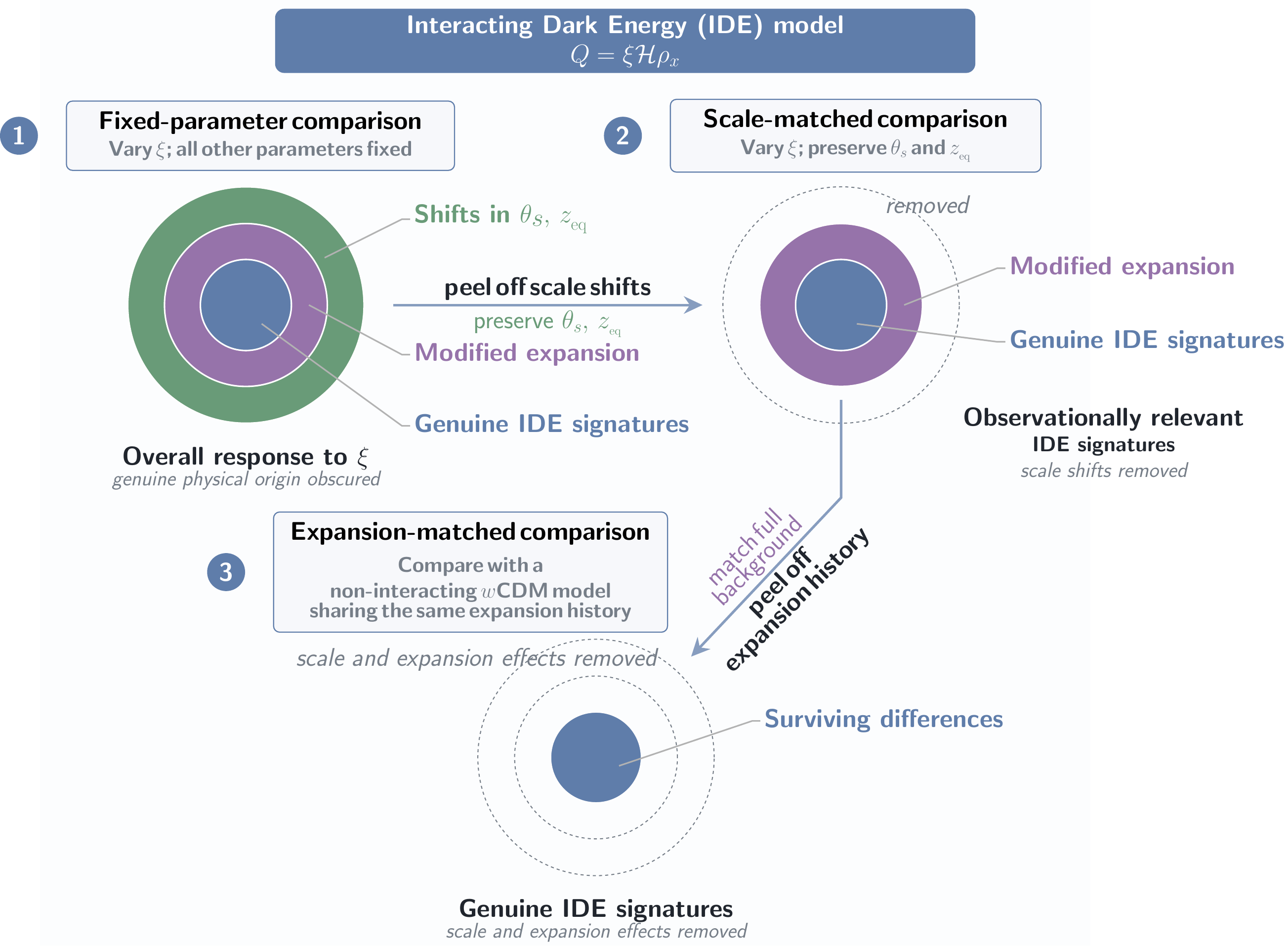}
\caption{Schematic illustration of our controlled comparison strategy, which we represent through the metaphor of peeling an onion, with each step removing one layer of the overall response of cosmological observables to DM-DE interactions. At the core of the onion, we reach signatures which cannot be attributed either to shifts in accurately measured physical scales or to differences in the background expansion history: these are the ones we interpret as genuine IDE signatures.}
\label{fig:onion}
\end{figure*}

Preserving $\theta_s$ and $z_{\text{eq}}$ does not, however, guarantee that the complete expansion history is unchanged. For this reason, differences observed in the scale-matched comparison contain a mixture of two effects: a modified background history and perturbative IDE signatures. This is not to say that the former are not IDE effects: they are, and they are precisely the signatures of IDE which are targeted by background probes such as BAO and SNeIa, and which have potentially been observed in DESI data~\cite{Giare:2024smz,Li:2024qso,Silva:2025hxw,Pan:2025qwy,Li:2026xaz}. However, these signatures cannot be distinguished from those of models with the same background evolution, e.g.\ an appropriately redefined $w$CDM model. This is why it is clearly of interest to isolate signatures which arise as a consequence of this modified background evolution from more genuine perturbative IDE signatures. To do so, we associate each scale-matched IDE model with a non-interacting, \textit{expansion-matched} $w$CDM model. From Eqs.~(\ref{eq:solutionrhoc},\ref{eq:solutionrhox}), it is relatively straightforward to show that the IDE background expansion history can be written \textit{exactly} as the sum of a matter-like contribution and a DE-like contribution with effective EoS $w_{x,\text{eff}}=w_x+\xi/3$. The relation between the expansion-matched (EM) and scale-matched (SM) parameters is then given by the following:
\begin{align}
w_x^{\text{EM}}&=w_x+\frac{\xi}{3}=w_{x,\text{eff}}\,,
\label{eq:expansionmatchingw}\\
\rho_{c,0}^{\text{EM}}&=\rho_{c,0}^{\text{SM}}+\frac{\xi}{3w_x+\xi}\rho_{x,0}^{\text{SM}}\,,
\label{eq:expansionmatchingc}\\
\rho_{x,0}^{\text{EM}}&=\frac{3w_x}{3w_x+\xi}\rho_{x,0}^{\text{SM}}\,.
\label{eq:expansionmatchingx}
\end{align}
The resulting $w$CDM$_{\text{EM}}^A$ model has exactly the same background expansion history as the IDE$_{\text{SM}}^A$ model. Differences between the two, identified in our expansion-matched analysis to be discussed in Sec.~\ref{sec:genuineeffects}, can therefore be considered genuine perturbation-level IDE effects, which can discriminate between IDE and $w$CDM models sharing the same background expansion history.

As stressed earlier, for model class $A$ we set $w_x=-0.999$, which corresponds \textit{de facto} to an interacting vacuum model while avoiding the singularity in the fluid perturbations at $w_x=-1$. Since we set $\xi=-0.3$, this corresponds to an expansion-matched $w$CDM cosmology with $w_x^{\text{EM}} \simeq -1.1$, i.e.\ a phantom model. We then carry out the same hierarchy of controlled comparisons for model class $B$, where we instead set $w_x=-0.8$, while still setting $\xi=-0.3$. This corresponds to $w_x^{\text{EM}}=-0.9$, i.e.\ a quintessence-like model. Studying model class $B$ allows us not only to assess how IDE signatures depend on the DE EoS at fixed interaction strength, but also to study expansion-matched background expansion histories on the opposite side of the phantom divide.

In summary, the five comparisons we carry out for our six reference models are the following:
\begin{enumerate}
\item $\Lambda$CDM versus IDE$_{\text{FP}}^A$ (fixed-parameter comparison; Sec.~\ref{subsec:fixedparameter})
\item $\Lambda$CDM versus IDE$_{\text{SM}}^A$ (scale-matched comparison; Sec.~\ref{subsec:scalematched})
\item IDE$_{\text{SM}}^A$ versus $w$CDM$_{\text{EM}}^A$ (expansion-matched comparison; Sec.~\ref{sec:genuineeffects})
\item $\Lambda$CDM versus IDE$_{\text{SM}}^B$ (scale-matched comparison; Sec.~\ref{subsec:scalematched})
\item IDE$_{\text{SM}}^B$ versus $w$CDM$_{\text{EM}}^B$ (expansion-matched comparison; Sec.~\ref{sec:genuineeffects})
\end{enumerate}
A schematic illustration of our strategy is provided in Fig.~\ref{fig:onion}. We represent this through the metaphor of peeling an onion, with each (matching) step removing one layer of the overall response of cosmological observables to DM-DE interactions: the fixed-parameter comparison captures the na\"{i}ve response to IDE, whereas preserving $\theta_s$ and $z_{\text{eq}}$ peels away the outer layer associated with shifts in these scales, and matching the complete expansion history removes the next and final layer. What remains once the core is reached are signatures of IDE which cannot be attributed either to shifts in accurately measured physical scales or to differences in the background expansion history. Repeating the last two steps for class $B$ allows us to determine how these signatures depend on the DE EoS at fixed interaction strength.

\section{Removing scale shifts}
\label{sec:scalematched}

We now proceed through the first two stages of our controlled comparisons. We begin with the (misleading) fixed-parameter comparison for illustrative purposes, and then move on to the scale-matched comparison.

\subsection{Fixed-parameter comparison}
\label{subsec:fixedparameter}

\begin{figure*}[!htbp]
\centering
\includegraphics[width=0.48\linewidth]{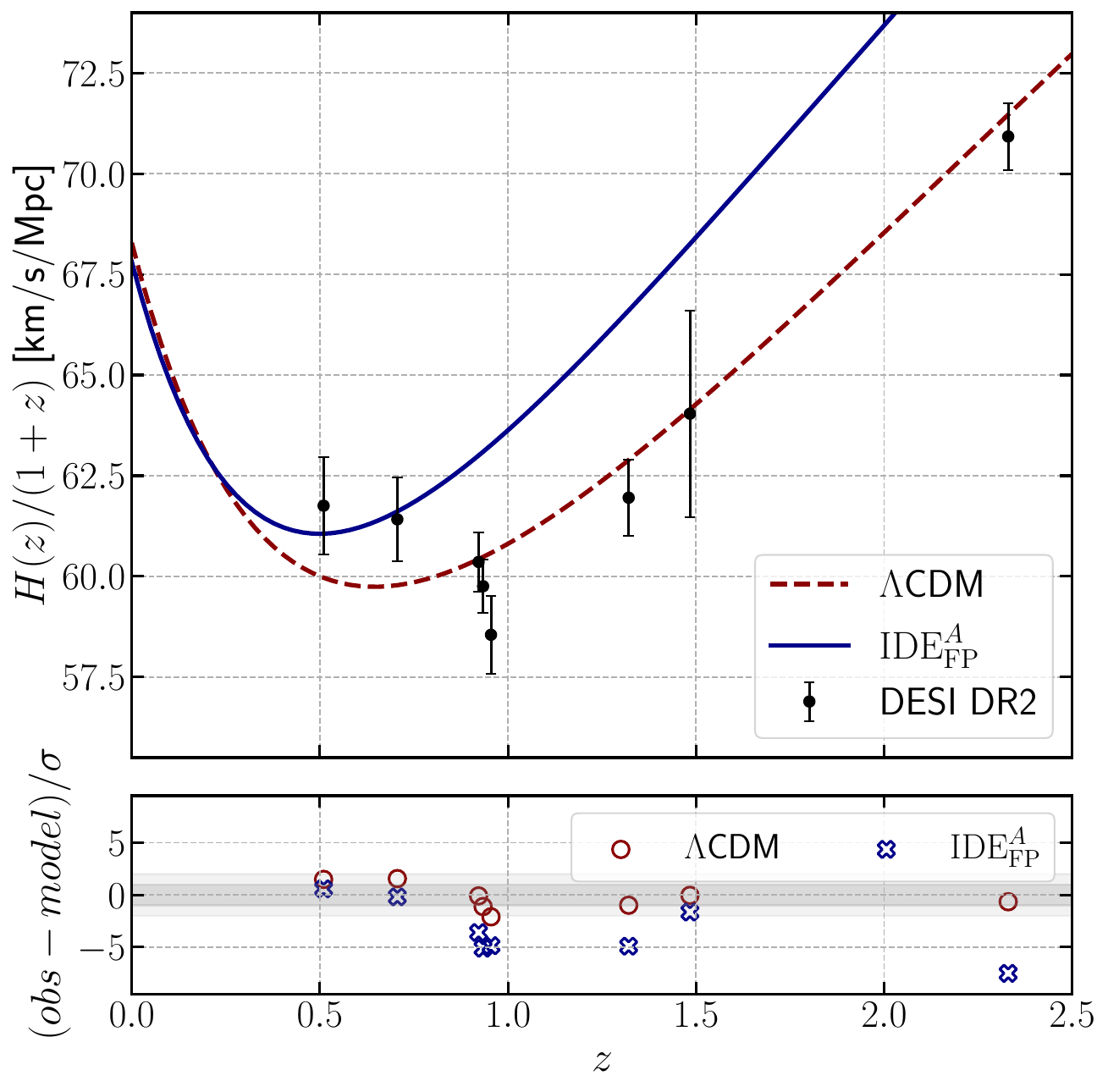} \quad \includegraphics[width=0.48\linewidth]{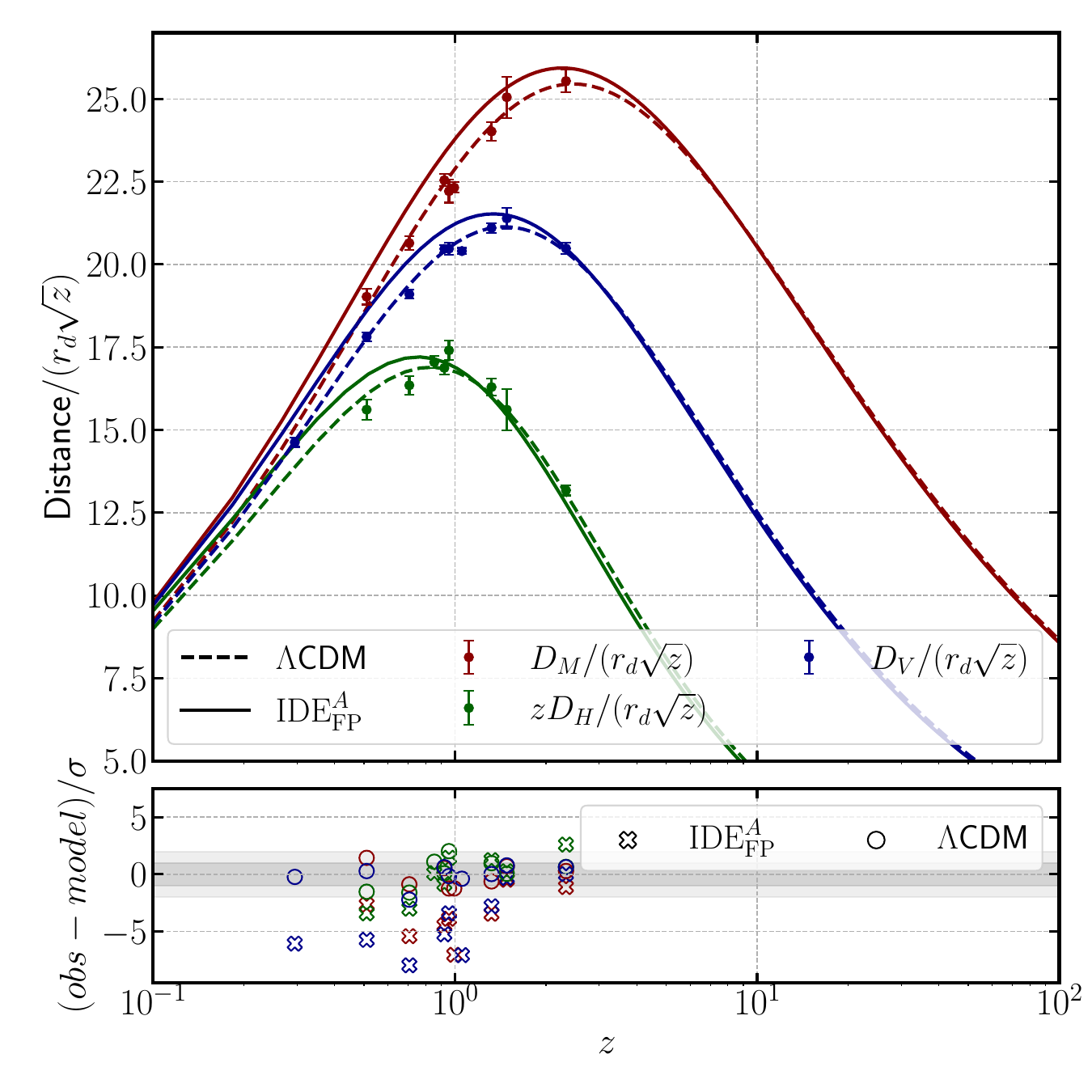}
\caption{Fixed-parameter comparison of background observables between the $\Lambda$CDM and IDE$_{\text{FP}}^A$ models, obtained by varying $\xi$ while keeping all other cosmological parameters (including $H_0$) fixed. \textit{Left panel}: appropriately rescaled expansion rate $H(z)/(1+z)$, with the red dashed and blue solid curves corresponding to $\Lambda$CDM and IDE$_{\text{FP}}^A$ respectively, and the black points corresponding to DESI DR2 line-of-sight BAO data. \textit{Right panel}: appropriately rescaled distances relevant for BAO measurements. The three types of distance measurements are distinguished by color, whereas the dashed and solid curves correspond to $\Lambda$CDM and IDE$_{\text{FP}}^A$ respectively, and the colored points correspond to DESI DR2 BAO measurements. In both cases, the lower sub-panels show normalized residuals, defined as the difference between each data point and the corresponding model prediction, normalized by the uncertainty of that data point. The residuals for the two cosmologies are distinguished by marker type. The dark gray and light gray bands indicate the $\pm1\sigma$ and $\pm2\sigma$ regions respectively. We stress that the DESI DR2 measurements are included only for illustrative purposes, to give a sense of the size of the shifts relative to current observational uncertainties: they are not fitted, nor should any conclusion regarding goodness of fit be drawn by eye.}
\label{fig:fpbackground}
\end{figure*}

We begin by examining the outermost layer of the onion shown in Fig.~\ref{fig:onion}, comparing the reference $\Lambda$CDM model against the IDE$_{\text{FP}}^A$ model. The effects of this fixed-parameter comparison can first be understood at the background level. We note that, in the early- vs late-time model dichotomy, the IDE model in question is typically classified as a late-time modification, the reason being that the instantaneous energy exchange rate $Q \propto \rho_x$ is negligible at high redshifts, where $\rho_x$ is negligible. However, as Eq.~(\ref{eq:solutionrhoc}) shows, this does \textit{not} imply that the early-time background is unchanged, even when $\rho_{c,0}$ is fixed. Since $\xi<0$ and energy is transferred from DM to DE, the DM density in the past had to be larger relative to $\Lambda$CDM in order to reach the same value of $\rho_{c,0}$ today. As we can observe in the left panel of Fig.~\ref{fig:fpbackground}, this enhances the expansion rate $H(z)$ over a large redshift range. The larger value of the DM density in the past also causes matter-radiation equality to occur earlier, as can be seen from the larger value of $z_{\text{eq}}$ in Tab.~\ref{tab:referencecosmologies}: this will have a huge impact on the CMB, as we shall discuss shortly. The enhancement in $H(z)$ has the effect of reducing both the comoving sound horizon at recombination $r_s^{\star}$ and the comoving angular diameter distance to the surface of last scattering $D_A^{\star}$. Since these changes act in opposite directions on the acoustic angular scale $\theta_s=r_s^{\star}/D_A^{\star}$, their net effect is hard to estimate analytically. Numerically, we find that, for the IDE$_{\text{FP}}^A$ model, the effect on $D_A^{\star}$ is the dominant one. As can be seen in Tab.~\ref{tab:referencecosmologies}, the net effect is therefore to increase $\theta_s$, with $100\theta_s$ increasing from $\simeq 1.04$ to $\simeq 1.06$. At this point, it is worth recalling that the acoustic angular scale is constrained to $100\theta_s=1.0411 \pm 0.0003$ by \textit{Planck}~\cite{Planck:2018vyg}. This makes the shift in question nominally excluded at $\gg 10\sigma$, clarifying why a fixed-parameter comparison of this type is misleading at best. The same shifts discussed above also affect BAO observables, as can be seen in the right panel of Fig.~\ref{fig:fpbackground}.

Another important scale affected by the shift in the epoch of matter-radiation equality is the comoving equality wavenumber, $k_{\text{eq}} \equiv a_{\text{eq}}H_{\text{eq}}$. At early times, the DM-DE interaction leads to an additional matter-like contribution, see Eq.~(\ref{eq:solutionrhoc}), so the physical matter density relevant at equality is $\omega_{m,\text{eff}} \equiv \omega_m+\xi\omega_x/(3w_{x,\text{eff}})$. Neglecting the DE-like contribution at equality and recalling that the radiation density is fixed, it is easy to show the of $k_{\text{eq}}$ in IDE relative to its $\Lambda$CDM value is given by the following:
\begin{equation}
\frac{k_{\text{eq}}^{\text{IDE}_{\text{FP}}^A}}{k_{\text{eq}}^{\Lambda\text{CDM}}}\simeq\frac{\omega_{m,\text{eff}}}{\omega_m}=1+\frac{\xi\omega_x}{3\omega_mw_{x,\text{eff}}}\,.
\label{eq:keq}
\end{equation}
As both $\xi$ and $w_{x,\text{eff}}$ are negative for IDE$_{\text{FP}}^A$, the interaction increases the effective early-time matter density. This makes equality occur earlier, at a smaller value of $a_{\text{eq}}$ and when the expansion rate was larger, implying that the corresponding comoving horizon was smaller and $k_{\text{eq}}$ is shifted towards larger wavenumbers, consistently with Eq.~(\ref{eq:keq}).

\begin{figure*}[!htbp]
\centering
\includegraphics[height=0.9\textwidth]{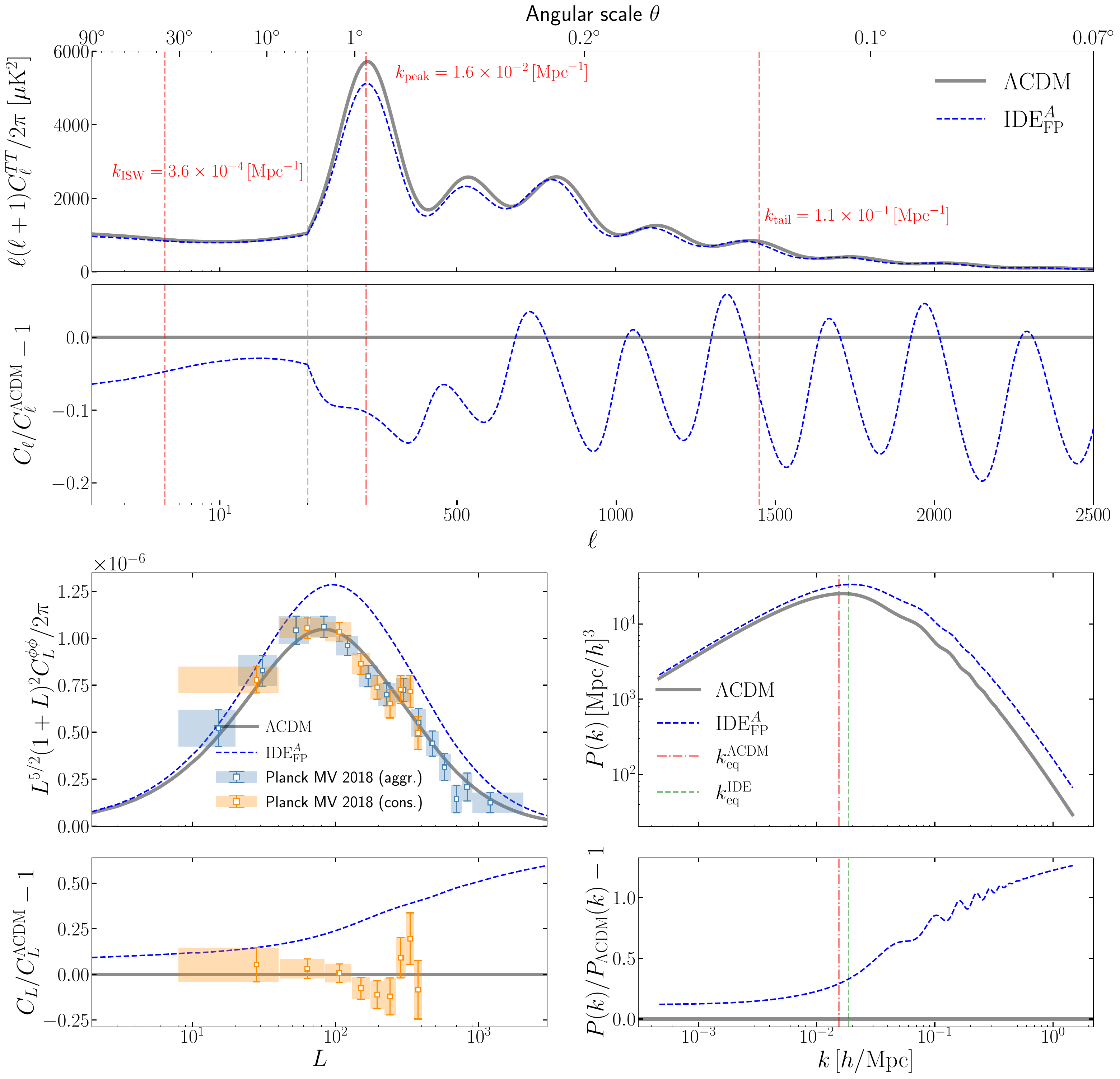}
\caption{Fixed-parameter comparison of the CMB temperature power spectrum, CMB lensing potential power spectrum, and linear matter power spectrum between the $\Lambda$CDM and IDE$_{\text{FP}}^A$ models, obtained by varying $\xi$ while keeping all other cosmological parameters (including $H_0$) fixed. In all relevant sub-panels, the gray solid and blue dashed curves correspond to $\Lambda$CDM and IDE$_{\text{FP}}^A$ respectively. \textit{Upper panel}: appropriately rescaled CMB temperature power spectrum, with the corresponding angular scale $\theta$ shown on the upper horizontal axis. The three red vertical lines indicate the multipoles associated with three representative Fourier modes we will study later: $k_{\text{ISW}}=3.6 \times 10^{-4}\,\text{Mpc}^{-1}$, $k_{\text{peak}}=1.6 \times 10^{-2}\,\text{Mpc}^{-1}$, and $k_{\text{tail}}=1.1 \times 10^{-1}\,\text{Mpc}^{-1}$, which mainly contribute to the late ISW signal, the first acoustic peak, and the damping tail respectively. The gray dashed vertical line indicates the transition between logarithmic and linear scales on the multipole axis. \textit{Lower left panel}: appropriately rescaled CMB lensing potential power spectrum, with the light blue and orange points and bands showing the Planck 2018 reconstruction obtained using the aggressive and conservative multipole selections respectively. These measurements are included only for illustrative purposes, to give a sense of the size of the shifts relative to current observational uncertainties: they are not fitted, nor should any conclusion regarding goodness of fit be drawn by eye. \textit{Lower right panel}: linear matter power spectrum, with the red dash-dotted and green dashed vertical lines corresponding to the equality wavenumbers in $\Lambda$CDM and IDE$_{\text{FP}}^A$ respectively. For each of the three observables, the lower sub-panels show the fractional difference relative to the corresponding $\Lambda$CDM prediction, with the conservative Planck lensing residuals also shown in the corresponding CMB lensing sub-panel.}
\label{fig:fp}
\end{figure*}

The resulting changes in the CMB temperature power spectrum, CMB lensing power spectrum, and matter power spectrum are instead shown in Fig.~\ref{fig:fp}. The effect of the increase in $\theta_s$ discussed earlier is to shift all acoustic peaks towards lower multipoles, as expected and as can be seen in the upper sub-panel of Fig.~\ref{fig:fp}. These shifts also explain the oscillatory patterns observed in the high-$\ell$ residuals. On the other hand, the earlier onset of matter domination has the effect of reducing the early integrated Sachs-Wolfe (ISW) effect. The reason is that, by the time of recombination, there is a significantly smaller radiation fraction (relative to matter), and therefore less evolution of the gravitational potentials which sources the early ISW effect. This reduced early ISW effect shows up as a very strong suppression of the first acoustic peak. There is also a much slighter reduction of power at very low multipoles: although the signal is in the cosmic variance-dominated regime, and hence of little interest, we note that it is perfectly compatible with the reduction in the late ISW effect observed in Ref.~\cite{Escamilla:2023oce} in the context of (effective) phantom cosmologies. The shift in the equality epoch also leaves a large imprint on the linear matter power spectrum (lower right sub-panel). In particular, the shift of $k_{\rm eq}$ towards larger wavenumbers discussed above explains the corresponding shift of the turnover in $P_m(k)$ to the right. The overall enhancement of $P_m(k)$ across the whole range is instead a consequence of the modified transfer function and growth history. Similar effects are responsible for the overall enhancement of the CMB lensing power spectrum $C_L^{\phi\phi}$, as well as the shift of the peak towards larger multipoles. We recall, in fact, that the CMB lensing power spectrum is sensitive to the projected comoving horizon at matter-radiation equality~\cite{Baxter:2020qlr}. Because the fixed-parameter comparison simultaneously affects physical scales, background expansion, and the evolution of perturbations, a more detailed discussion of the origin of these effects would be ambiguous. It is only once we start peeling the outer layers of the onion that a physical interpretation of this type becomes more meaningful. For this reason, a detailed study of the physical origin of shifts in cosmological observables is deferred to the scale- and expansion-matched comparisons.

\subsection{Scale-matched comparison}
\label{subsec:scalematched}

Having examined the outermost layer of the onion in Fig.~\ref{fig:onion}, we now peel it away, keeping $\theta_s$ and $z_{\text{eq}}$ fixed in order to address the shortcomings observed in the fixed-parameter comparison. We therefore set up the scale-matched models IDE$_{\text{SM}}^A$ and IDE$_{\text{SM}}^B$, which by construction share the same values of $\theta_s$ and $z_{\text{eq}}$ as the reference $\Lambda$CDM model. In practice, as far as $\theta_s$ is concerned, we fix it by supplying its value as an input parameter to \texttt{CLASS} (recall that in the previous fixed-parameter comparison we instead supplied $H_0$ as an input parameter). We then lower $\omega_c$ to match the reference value of $z_{\text{eq}}$ (see Tab.~\ref{tab:referencecosmologies}). The fact that we need to lower $\omega_c$ should not come as a surprise, as it is a consequence of the extra matter-like contribution generated by the DM-DE interaction [see Eq.~(\ref{eq:solutionrhoc})]. The resulting shifts in $H_0$ and $\Omega_m$ observed in Tab.~\ref{tab:referencecosmologies} are a consequence of this matching procedure: since we are supplying $\theta_s$, \texttt{CLASS} internally derives the value of $H_0$ required to match the desired value of $\theta_s$. This leads to an increase in $H_0$ to $\sim 71.3\,{\text{km}}/{\text{s}}/{\text{Mpc}}$, consistent with the fact that these models have long been of interest in the context of the Hubble tension~\cite{DiValentino:2019ffd,DiValentino:2019jae,Zhai:2023yny}, and a significant decrease in $\Omega_m$ to $\sim 0.21$, itself a consequence of both the increase in $H_0$ and the decrease in $\omega_c$. We recall that the purpose of this scale-matched comparison is to remove the dominant effects observed in Fig.~\ref{fig:fp}, which are associated with shifts in $\theta_s$ and $z_{\text{eq}}$, and work out the interaction signatures which survive once these two accurately measured scales are preserved.

\begin{figure*}[!htbp]
\centering
\includegraphics[width=0.47\linewidth]{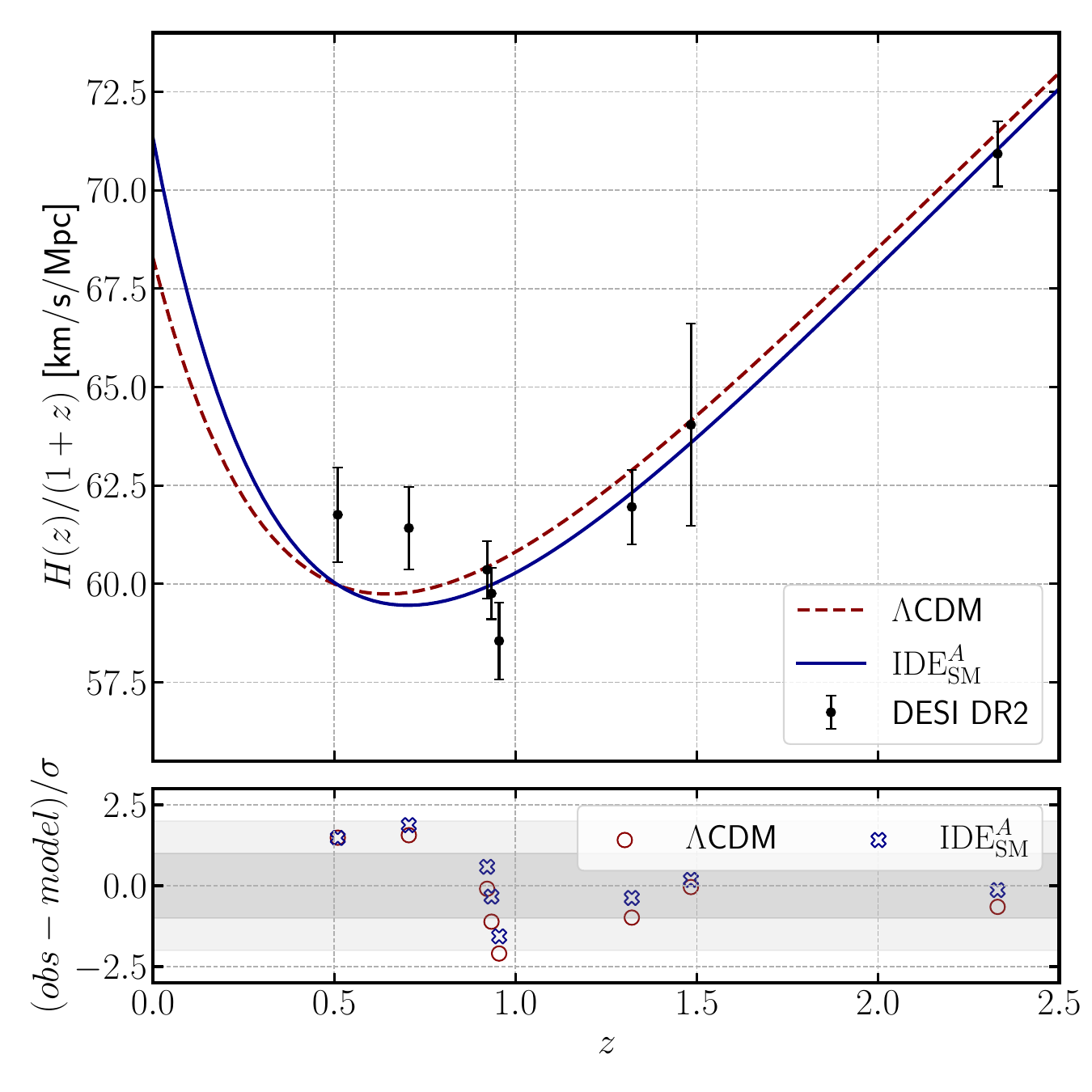} \quad \includegraphics[width=0.47\linewidth]{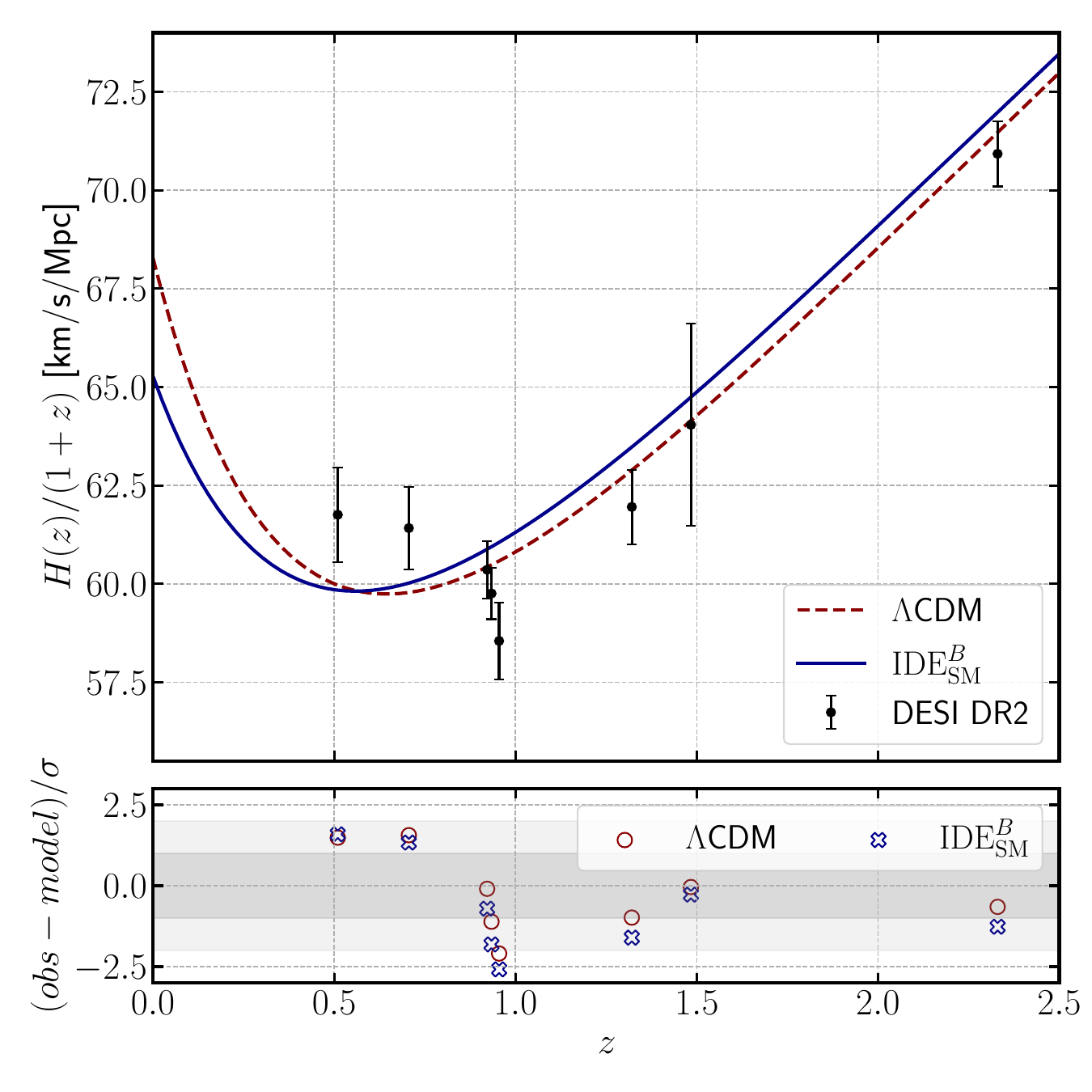} \\
\includegraphics[width=0.47\linewidth]{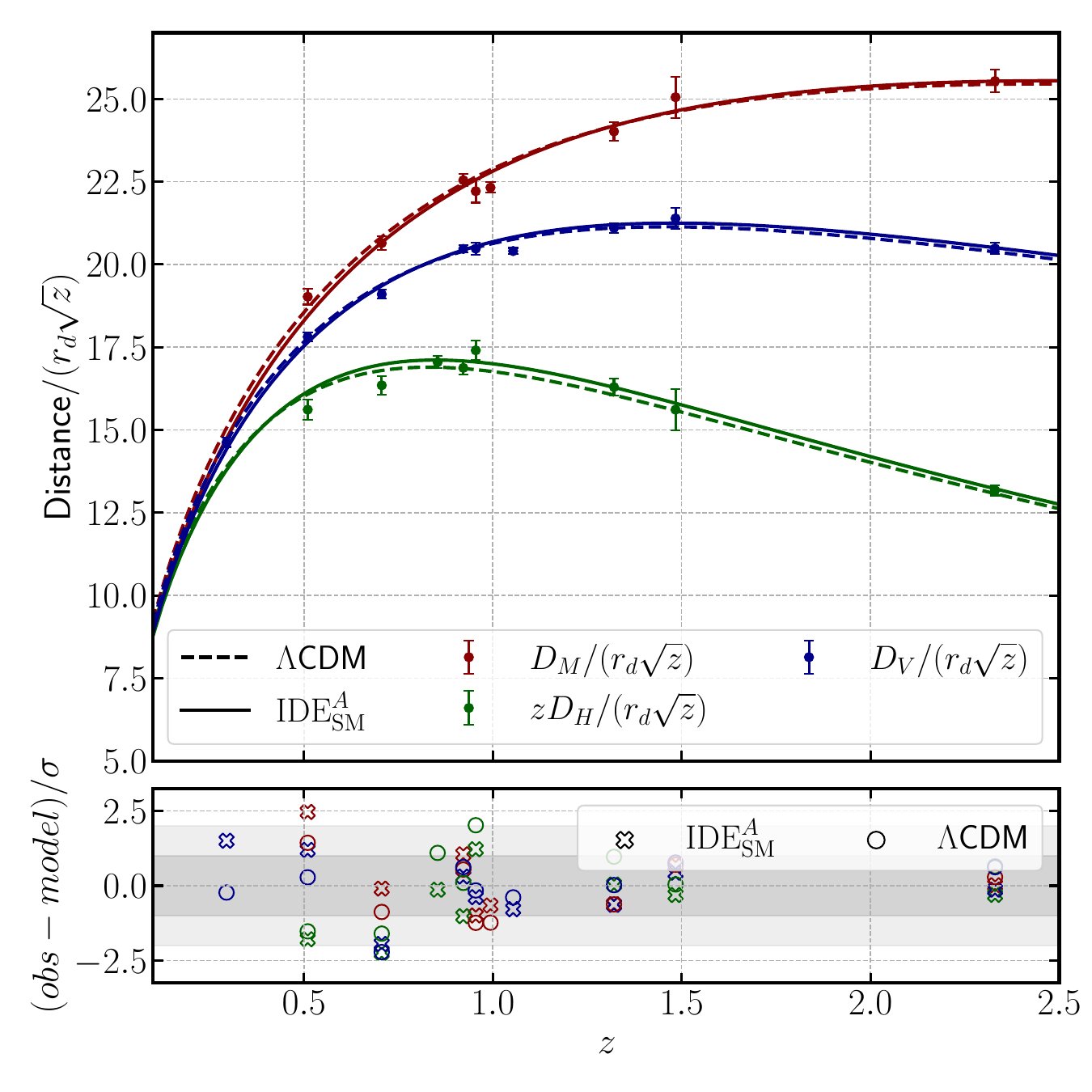} \quad \includegraphics[width=0.47\linewidth]{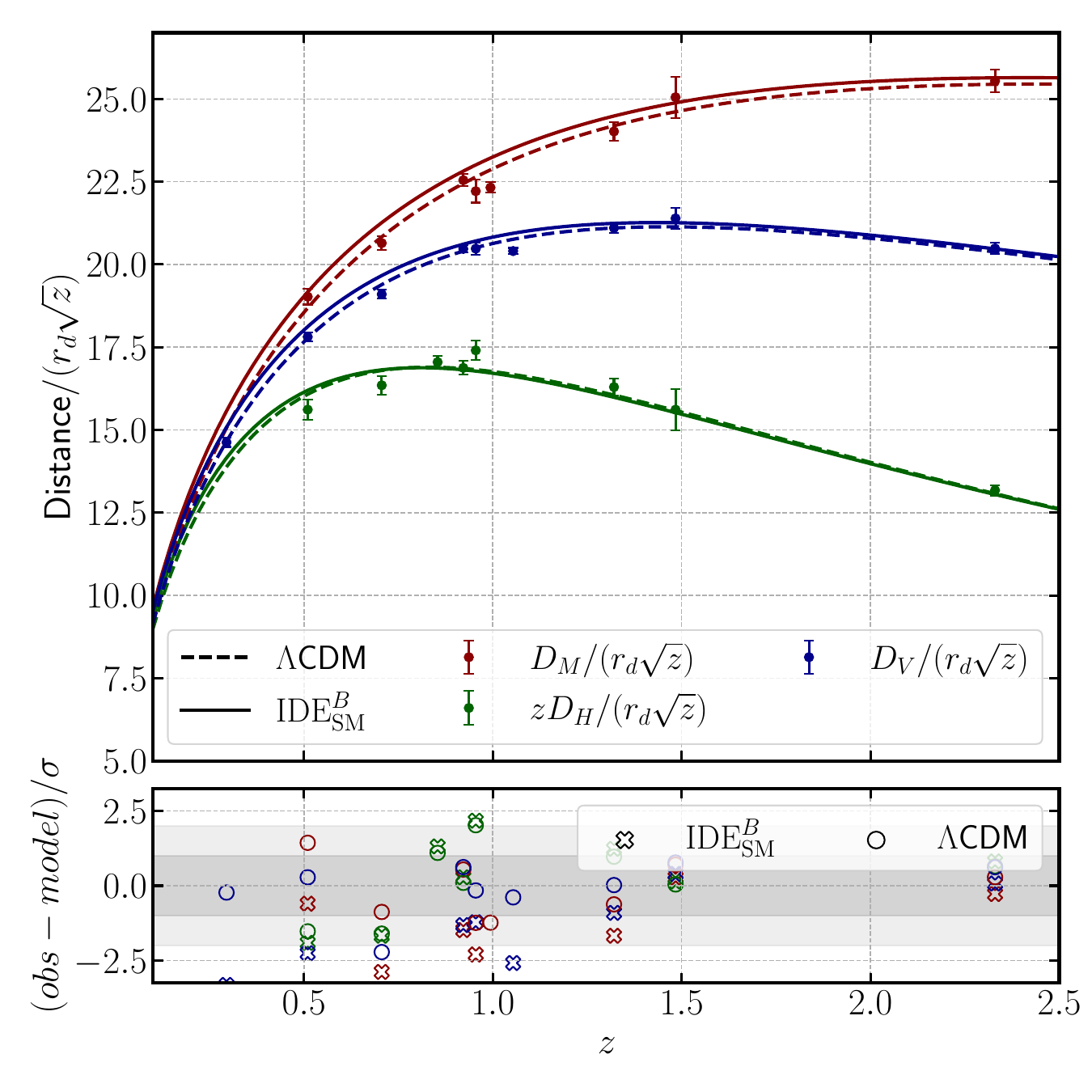}
\caption{As in Fig.~\ref{fig:fpbackground}, but for the scale-matched comparisons between $\Lambda$CDM and IDE$_{\text{SM}}^A$ (left column) and between $\Lambda$CDM and IDE$_{\text{SM}}^B$ (right column), with the two model classes distinguished by the equation of state of their DE components. The upper and lower rows show the appropriately rescaled expansion rate and BAO distances respectively. For each model class, we carry out the scale matching by adjusting $\omega_c$ to preserve $z_{\text{eq}}$, while $\theta_s$ is fixed as an input parameter, so $H_0$ changes as a derived parameter. We retain all color, data, and marker conventions from Fig.~\ref{fig:fpbackground} and, as there, the DESI DR2 measurements are included only for illustrative purposes.}
\label{fig:smbackground}
\end{figure*}

The above shifts, besides preserving $\theta_s$ and $z_{\text{eq}}$, essentially restore the early-time background evolution to be close to that of the reference $\Lambda$CDM model (since the DE component is completely negligible at high redshifts). However, the late-time expansion history can still deviate significantly from $\Lambda$CDM. This is clearly illustrated in Fig.~\ref{fig:smbackground}. For model class $A$ (upper left sub-panel), the effective phantom EoS $w_{x,\text{eff}}^A \simeq -1.1$ leads to a larger present-day expansion rate $H_0$, as is well known given the degeneracy between $H_0$ and the DE EoS~\cite{Vagnozzi:2019ezj,Alestas:2020mvb}. At intermediate redshifts, the expansion rate then drops below the $\Lambda$CDM one. As expected, the opposite behavior is observed for model class $B$ (upper right sub-panel), which displays an effective quintessence-like behavior, with $w_{x,\text{eff}}^B \simeq -0.9$: in this case, both the present-day and the late-time expansion rates are lower relative to $\Lambda$CDM, before increasing at intermediate redshifts. These changes are required since preserving $\theta_s$ when the sound horizon is nearly unchanged requires the distance to the surface of last scattering to be preserved as well. In the two lower sub-panels of Fig.~\ref{fig:smbackground}, we see how these differences in the expansion rate propagate into the transverse, radial, and volume-averaged distances relevant for BAO measurements. For classes $A$ (lower left sub-panel) and $B$ (lower right sub-panel), $D_M$ and $D_V$ shift in opposite directions, whereas the shifts in $D_H$ directly reflect those in $H(z)$ observed above. We only include the DESI measurements to allow the reader to gauge by eye the size of these shifts relative to current observational uncertainties, and caution against using them to carry out visual goodness of fit analyses. In summary, matching $\theta_s$ and $z_{\text{eq}}$ alone still leaves non-negligible differences in the late-time background geometry, i.e.\ in the late-time expansion rate and distances: it is precisely these signatures which are being constrained by current BAO and SNeIa data.

\begin{figure*}[!htbp]
\centering
\includegraphics[height=0.9\textwidth]{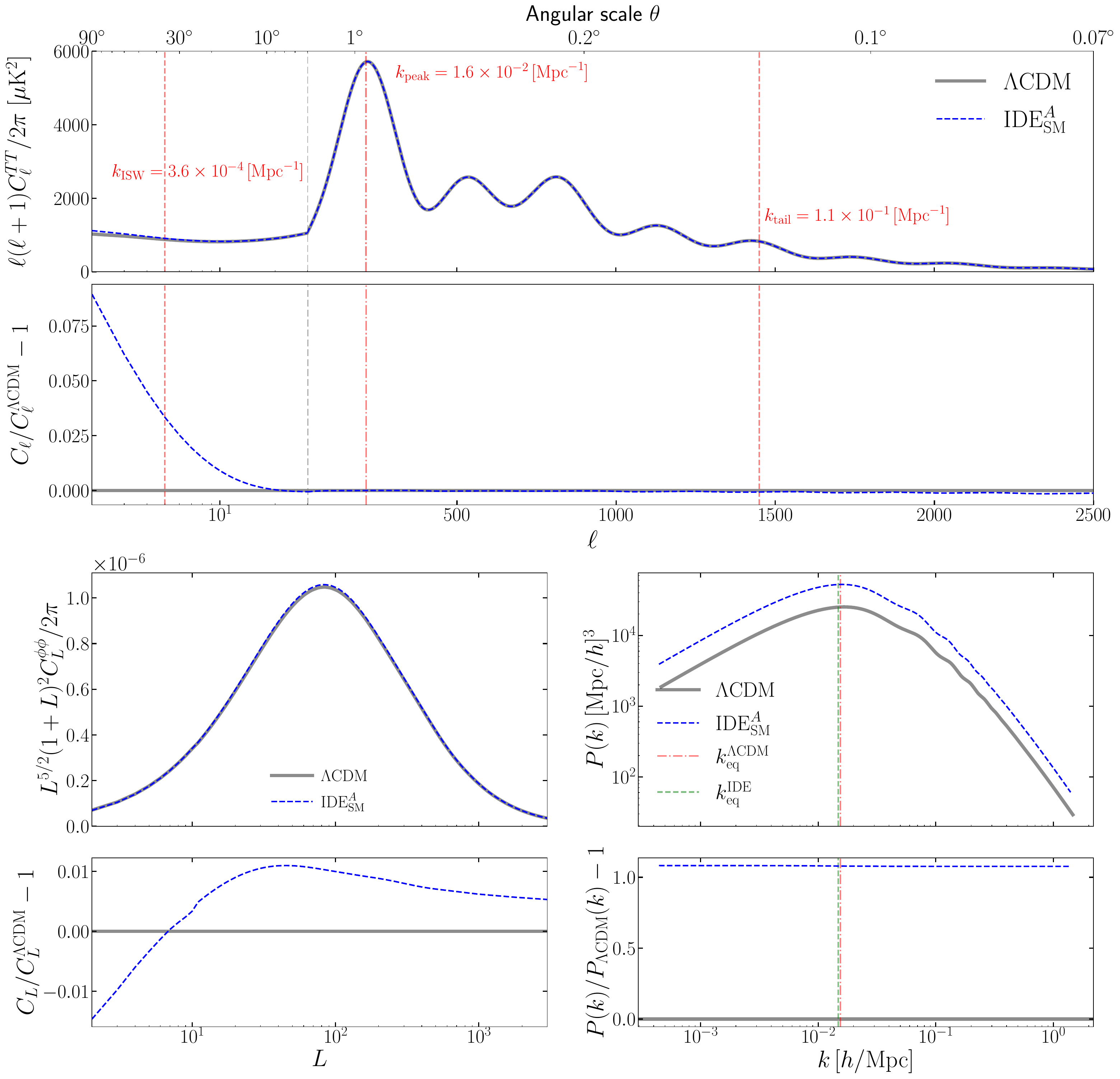}
\caption{As in Fig.~\ref{fig:fp}, but for the scale-matched comparison between $\Lambda$CDM and IDE$_{\text{SM}}^A$. In all relevant sub-panels, the gray solid and blue dashed curves correspond to $\Lambda$CDM and IDE$_{\text{SM}}^A$ respectively. The slight separation between the two values of $k_{\text{eq}}$ in the matter power spectrum sub-panel arises because $k$ is plotted in units of $h\,{\text{Mpc}}^{-1}$ and the two models, while having the same value of $z_{\text{eq}}$, have different values of $H_0$; when expressed in ${\text{Mpc}}^{-1}$, the two values of $k_{\text{eq}}$ are identical by construction.}
\label{fig:sma}
\end{figure*}

\begin{figure*}[!htbp]
\centering
\includegraphics[height=0.9\textwidth]{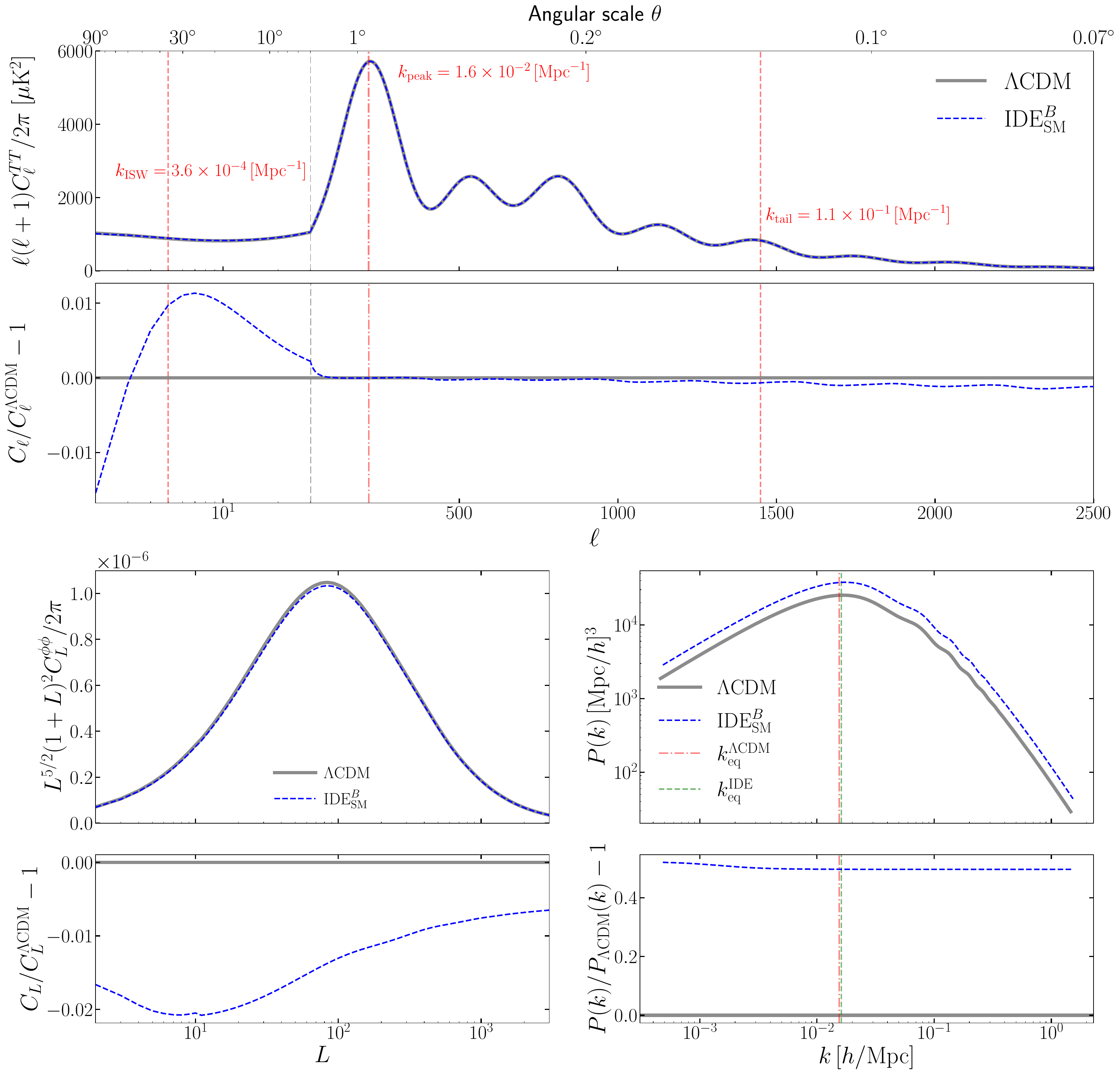}
\caption{As in Fig.~\ref{fig:sma}, but for the scale-matched comparison between $\Lambda$CDM and IDE$_{\text{SM}}^B$.}
\label{fig:smb}
\end{figure*}

We show the impact of this scale-matching procedure on the main cosmological observables in Fig.~\ref{fig:sma} (for model class $A$) and Fig.~\ref{fig:smb} (for model class $B$), which are the scale-matched counterparts of Fig.~\ref{fig:fp}. The most striking result concerns the CMB temperature power spectrum (two upper sub-panels). In contrast to the fixed-parameter comparison (upper sub-panel of Fig.~\ref{fig:fp}), the acoustic peaks in both the IDE$_{\text{SM}}^A$ and IDE$_{\text{SM}}^B$ models essentially coincide with those of the reference $\Lambda$CDM model. This occurs because matching $\theta_s$ restores their positions, while matching $z_{\text{eq}}$ restores their (absolute and relative) amplitudes, since the baryonic and primordial power spectrum parameters have also been left unchanged. For both model classes, we find residual differences on intermediate and small scales ($\ell \gtrsim 100$) which are well below the per-mille level. More significant differences can be observed on the largest angular scales, $\ell \lesssim 30$: these reflect the different late-time evolution of the gravitational potentials, and therefore a different late ISW imprint. The direction and magnitude of these low-$\ell$ shifts are model-dependent, but in both cases are consistent with a modified late ISW contribution (see for instance Fig.~1 of Ref.~\cite{Escamilla:2023oce}). We nonetheless stress that these differences are still small, at the percent level, and lie within the cosmic variance-dominated regime, therefore carrying very limited constraining power. In short, we find that matching $\theta_s$ and $z_{\text{eq}}$ within this IDE model is sufficient to produce nearly indistinguishable CMB temperature power spectra, in contrast to the huge changes observed in na\"{i}ve fixed-parameter comparisons.

These results do not extend to the CMB lensing and matter power spectra, shown in the lower sub-panels of Fig.~\ref{fig:sma} and Fig.~\ref{fig:smb}. The main reason is that, while CMB temperature anisotropies are mainly determined by the physics around recombination, CMB lensing and matter clustering are sensitive to the subsequent evolution of the gravitational potentials and matter perturbations. As we have clearly seen in Fig.~\ref{fig:smbackground}, matching $\theta_s$ and $z_{\text{eq}}$ is not sufficient to match the late-time expansion rate. This on its own is sufficient to modify the growth history, which can therefore lead to changes in the CMB lensing and matter power spectra (especially in the latter). These changes therefore still combine the effects of a different late-time background, a different matter content, and the different evolution of perturbations due to the interaction. For this reason, it is wise to postpone a detailed physical interpretation to the expansion-matched comparison, where the changes in the background can be removed, and the genuine interaction effects can therefore be dissected more cleanly. We clarify a subtle point regarding the $P_m(k)$ plots (lower right sub-panels of Fig.~\ref{fig:sma} and Fig.~\ref{fig:smb}), where one immediately notices that the two markers for $k_{\text{eq}}$ within the $\Lambda$CDM and IDE models appear slightly displaced, despite $z_{\text{eq}}$ having been fixed. We stress that this is simply due to the conventional choice of expressing wavenumber in units of $h\,{\text{Mpc}}^{-1}$, and therefore reflects the different values of $H_0$ in $\Lambda$CDM versus IDE (see Tab.~\ref{tab:referencecosmologies}): had we chosen to report these in units of ${\text{Mpc}}^{-1}$ instead, the two markers would have coincided.

\begin{figure*}[!htbp]
\centering
\includegraphics[width=1\linewidth]{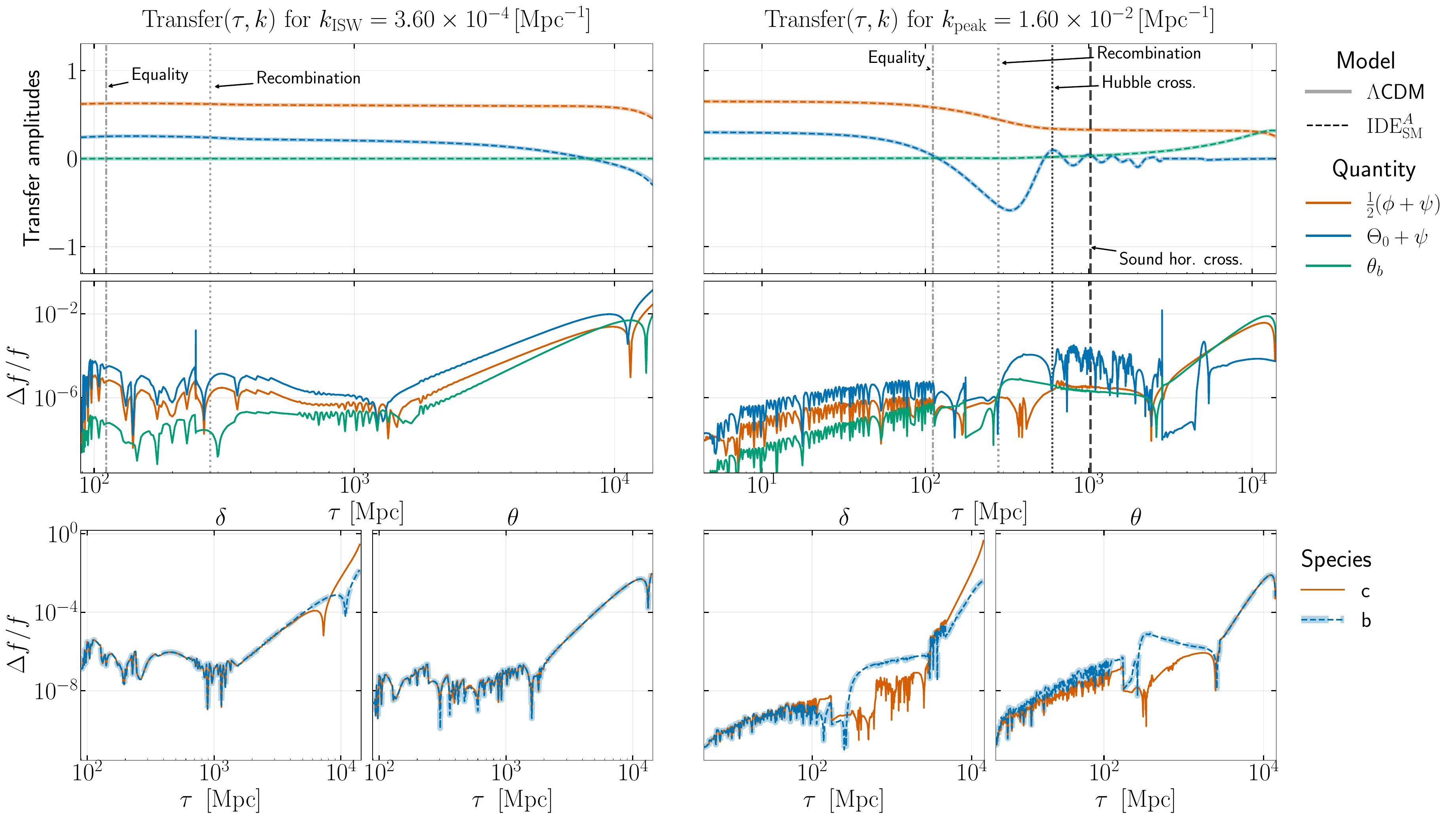} 
\caption{Evolution of the relevant transfer functions and matter perturbations in the scale-matched comparison between $\Lambda$CDM and IDE$_{\text{SM}}^A$, shown as functions of conformal time $\tau$, for the representative modes $k_{\text{ISW}}=3.6 \times 10^{-4}\,\text{Mpc}^{-1}$ (left column) and $k_{\text{peak}}=1.6 \times 10^{-2}\,\text{Mpc}^{-1}$ (right column). \textit{Upper row}: evolution of the Weyl potential $(\phi+\psi)/2$ (orange), effective (gravitationally redshifted) photon monopole $\Theta_0+\psi$ (blue), and baryon velocity divergence $\theta_b$ (green). In the upper row, solid and dashed curves correspond to $\Lambda$CDM and IDE$_{\text{SM}}^A$ respectively, while the lower sub-panels show their appropriately normalized differences (defined later in the caption). The vertical lines mark matter-radiation equality and recombination, as well as Hubble crossing and sound horizon crossing where relevant. \textit{Lower row}: for each representative mode, the left and right sub-panels show the appropriately normalized differences between the density contrasts $\delta$ and velocity divergences $\theta$ of IDE$_{\text{SM}}^A$ and $\Lambda$CDM. The orange solid and blue dashed curves correspond to DM ($c$) and baryons ($b$) respectively. In all residual sub-panels, the quantity denoted by $\Delta f/f$ is the absolute value of the difference between the IDE$_{\text{SM}}^A$ and $\Lambda$CDM predictions for the relevant function $f$, normalized by the maximum absolute value reached by the corresponding $\Lambda$CDM quantity across the whole range of $\tau$ considered, i.e.\ $(\Delta f/f)(k,\tau)\equiv \vert f^{\text{IDE}_{\text{SM}}^A}(k,\tau)-f^{\Lambda\text{CDM}}(k,\tau)\vert/\max_{\tau'}\vert f^{\Lambda\text{CDM}}(k,\tau')\vert$, with $f$ being the relevant transfer function or matter perturbation.}
\label{fig:smanatomytransfer}
\end{figure*}

We now seek to understand why this scale-matching procedure is sufficient to keep the CMB temperature power spectrum virtually unchanged. To do so, we temporarily move to Newtonian gauge for clarity and inspect the evolution of the relevant transfer functions (as a function of time and for a given wavenumber) in Fig.~\ref{fig:smanatomytransfer}, focusing solely on model class $A$ for simplicity. We do so for two representative modes: $k_{\text{ISW}}=3.6 \times 10^{-4}\,{\text{Mpc}}^{-1}$ (left sub-panels), which mainly contributes to the late ISW signal, and $k_{\text{peak}}=1.6 \times 10^{-2}\,{\text{Mpc}}^{-1}$ (right sub-panels), whose dominant contribution is around the first acoustic peak. We show the transfer functions for the Weyl potential $(\phi+\psi)/2$, the effective (gravitationally red-/blue-shifted) photon monopole $\Theta_0+\psi$, and the baryon velocity divergence $\theta_b$: these respectively trace the gravitational driving, effective temperature, and Doppler contributions to the primary CMB temperature anisotropies. We begin by noting that, since the radiation and baryon physical densities are fixed, preserving $z_{\text{eq}}$ causes the early-time expansion rate and gravitational potential evolution to basically coincide with their $\Lambda$CDM counterparts. This is true even though $\omega_c$ decreases, since what matters is the coefficient of the matter-like contribution to the early-time DM density, i.e.\ $\omega_{c,\text{early}}\equiv\omega_c+\xi\omega_x/(3w_x+\xi)$; see Eq.~(\ref{eq:solutionrhoc}). In fact, the reduction in $\omega_c$ is required \textit{precisely} to compensate for this additional matter-like contribution. In addition, preserving $\theta_s$ implies that the same acoustic phases project onto the same angular scales: this is not relevant for the $k_{\text{ISW}}$ mode, but is especially important for the $k_{\text{peak}}$ mode, as we can appreciate by looking at the upper right sub-panel of Fig.~\ref{fig:smanatomytransfer}.

Taken together, these considerations explain why the relevant transfer functions are virtually unchanged relative to their $\Lambda$CDM counterparts: in the upper sub-panels of Fig.~\ref{fig:smanatomytransfer}, the differences are too small to be appreciated by the naked eye, and the appropriately normalized residual sub-panels confirm that the changes are well below the per-mille level. These differences are slightly larger, but still extremely small, for the large-scale $k_{\text{ISW}}$ mode, especially after recombination. This is unsurprising, as DE starts becoming non-negligible at sufficiently late times, therefore impacting the evolution of the Weyl potential. This explains why the changes observed earlier in Fig.~\ref{fig:sma} and Fig.~\ref{fig:smb} are mainly concentrated at low multipoles, as the different evolution of the Weyl potential modifies the late ISW contribution to the CMB temperature power spectrum.

A complementary view of the above considerations is provided in the lower panel of Fig.~\ref{fig:smanatomytransfer}. There we see that the evolution of the DM and baryon density perturbations and velocity divergences is essentially identical up to recombination. The differences in growth only become more appreciable at later times, which are irrelevant for the CMB temperature power spectrum, but can and indeed do affect the matter power spectrum. In short, we conclude that once the scale-matching procedure is performed, the interaction leaves no substantial imprint on the perturbation dynamics of DM, baryons, and gravitational potentials around the time of recombination, while still modifying the late-time background expansion, and with it the evolution of gravitational potentials and matter perturbations. Although in Fig.~\ref{fig:smanatomytransfer} we have focused on model class $A$, identical considerations apply to model class $B$, explaining why the CMB temperature power spectrum is nearly identical.

\begin{figure*}[htbp]
\centering
\includegraphics[width=0.8\textwidth]{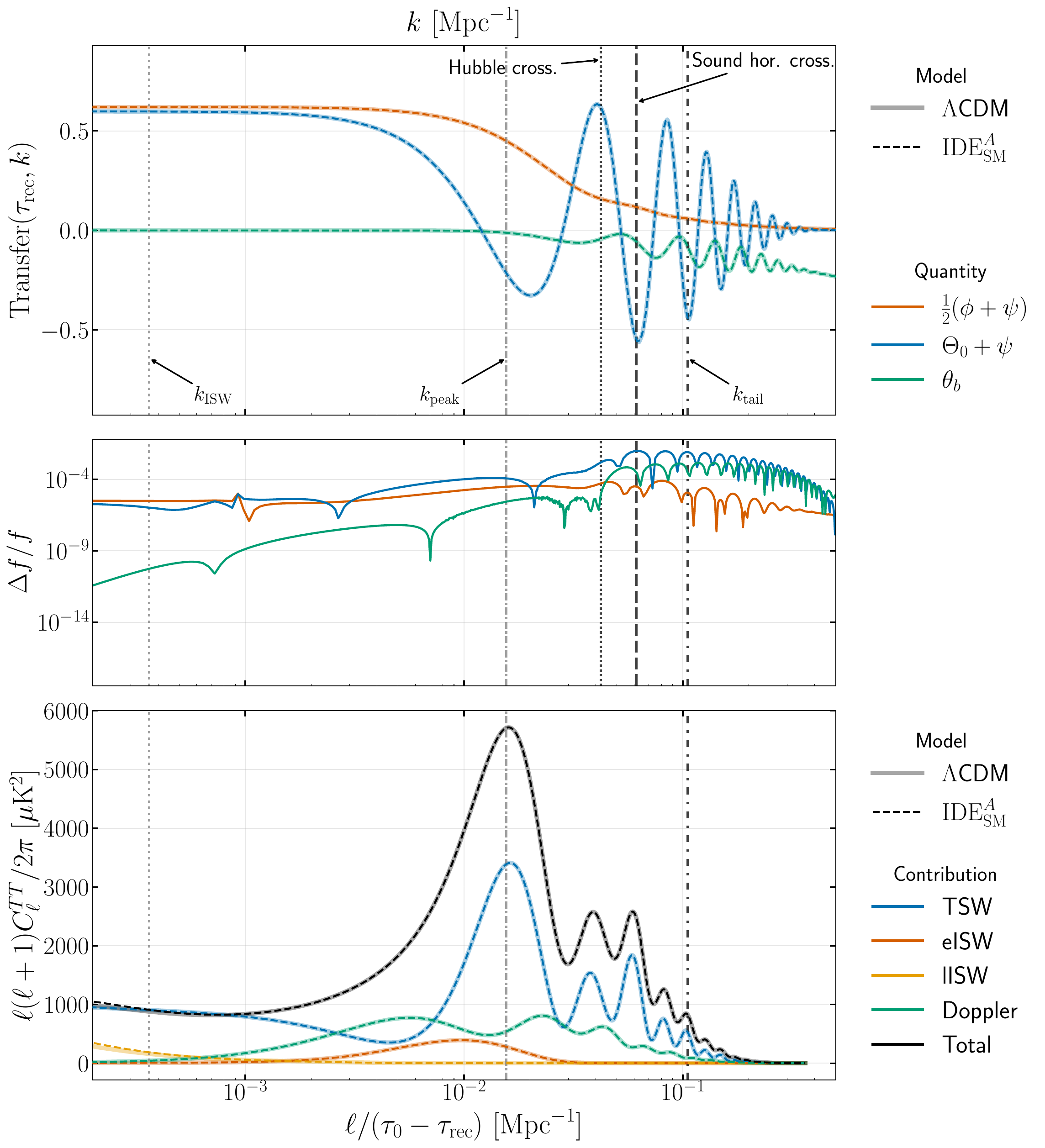}
\caption{Scale-matched comparison between $\Lambda$CDM and IDE$_{\text{SM}}^A$, illustrating the physical origin of the CMB temperature power spectrum. In all relevant sub-panels, the solid and dashed curves correspond to $\Lambda$CDM and IDE$_{\text{SM}}^A$ respectively. \textit{Upper panel}: Weyl potential $(\phi+\psi)/2$ (orange), effective photon monopole $\Theta_0+\psi$ (blue), and baryon velocity divergence $\theta_b$ (green), evaluated at recombination as a function of wavenumber $k$. The five vertical lines indicate the representative modes $k_{\text{ISW}}=3.6 \times 10^{-4}\,\text{Mpc}^{-1}$, $k_{\text{peak}}=1.6 \times 10^{-2}\,\text{Mpc}^{-1}$, and $k_{\text{tail}}=1.1 \times 10^{-1}\,\text{Mpc}^{-1}$, as well as the Hubble crossing and sound horizon-crossing scales. \textit{Middle panel}: appropriately normalized differences between the transfer functions, following the convention of Fig.~\ref{fig:smanatomytransfer}, but with the maximum taken over the plotted range of $k$ at recombination. \textit{Lower panel}: Sachs-Wolfe (TSW, blue), early ISW (eISW, dark orange), late ISW (lISW, light orange), and Doppler (light green) contributions to the CMB temperature power spectrum, shown alongside the total power spectrum. In order to facilitate comparison with the two upper panels, the horizontal axis is expressed in terms of the projection relation $k \simeq \ell/(\tau_0-\tau_{\text{rec}})$, which approximately maps each multipole to its characteristic wavenumber.}
\label{fig:smanatomycmb}
\end{figure*}

Yet another complementary view of the above results is shown in Fig.~\ref{fig:smanatomycmb}, whose upper sub-panel shows a snapshot of the same transfer functions, this time as a function of wavenumber instead of time, frozen at recombination. We see that the transfer functions at recombination are virtually identical to their $\Lambda$CDM counterparts across the whole range of modes relevant for the CMB, even for significantly larger wavenumbers than those studied earlier, for instance the mode labeled $k_{\text{tail}}$, which mainly contributes to the damping tail. The lower panels confirm these conclusions at the level of the different physical contributions to the CMB temperature power spectrum. In particular, we see that the Sachs-Wolfe and Doppler contributions are essentially identical to their $\Lambda$CDM counterparts. This is to be expected, given the agreement between their respective sources, i.e.\ the effective photon monopole $\Theta_0+\psi$ and the baryon velocity divergence $\theta_b$. The identical evolution of the Weyl potential around recombination instead explains why the eISW contribution is also unchanged. Again, the only relevant differences can be observed in the late ISW contribution at very low multipoles. However, as already stressed earlier, these differences are still very small and, importantly, appear in a region which is completely cosmic variance-dominated. Therefore, they are not expected to carry significant constraining power. In short, the above discussion and the transfer functions shown in Fig.~\ref{fig:smanatomytransfer} and Fig.~\ref{fig:smanatomycmb} explain why, once the acoustic and equality scales are matched, the CMB temperature power spectrum is essentially unaffected by the interaction between DM and DE: this is in sharp contrast to the significant changes observed in the fixed-parameter analysis (see Fig.~\ref{fig:fp}), whose misleading nature the reader will by now hopefully have no doubts about.

We close this comparison layer by briefly returning to the CMB lensing and matter power spectra (lower sub-panels of Fig.~\ref{fig:sma} and Fig.~\ref{fig:smb}). As stressed earlier, these differences, relatively small in the CMB lensing case but more significant for the matter power spectrum, still combine the effects of a different late-time background, a different matter content, and the different evolution of perturbations due to the interaction. As such, it is premature to interpret them as genuine interaction signatures. These considerations naturally set the stage for the expansion-matched comparison, where we compare each scale-matched IDE model against a non-interacting $w$CDM model tailored to have exactly the same background expansion history.

\section{Isolating genuine interaction effects}
\label{sec:genuineeffects}

\begin{figure*}[!htbp]
\centering
\includegraphics[height=0.9\textwidth]{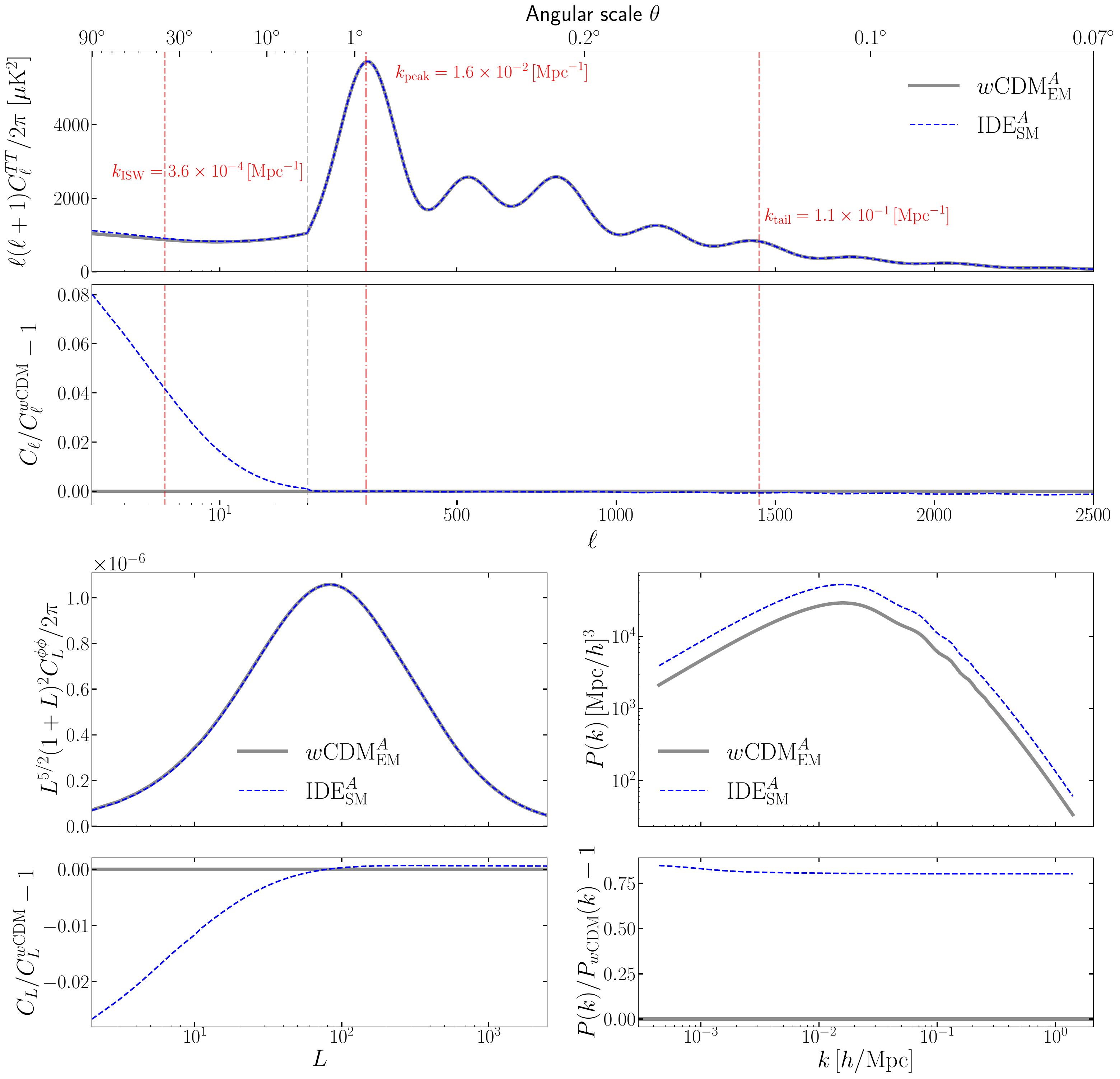}
\caption{As in Fig.~\ref{fig:sma}, but for the expansion-matched comparison between IDE$_{\text{SM}}^A$ and $w$CDM$_{\text{EM}}^A$.}
\label{fig:ema}
\end{figure*}

\begin{figure*}[!htbp]
\centering
\includegraphics[height=0.9\textwidth]{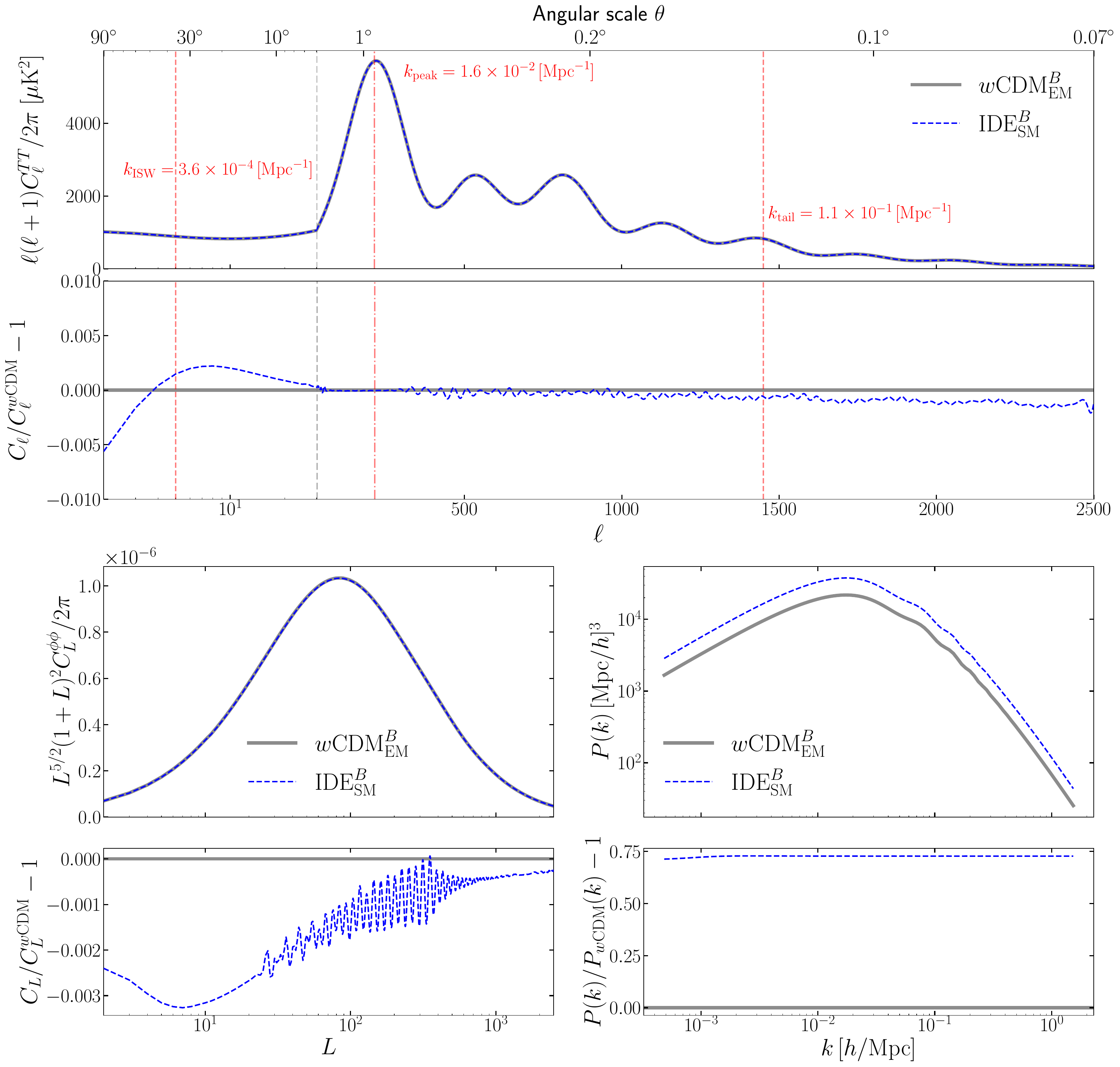}
\caption{As in Fig.~\ref{fig:ema}, but for the expansion-matched comparison between IDE$_{\text{SM}}^B$ and $w$CDM$_{\text{EM}}^B$.}
\label{fig:emb}
\end{figure*}

We are now ready to peel away the final layer of the onion in Fig.~\ref{fig:onion} in order to reach the core, where the genuine IDE signatures lie. We therefore turn to the expansion-matched comparison, where we compare each scale-matched IDE model to its non-interacting $w$CDM counterpart, constructed so as to match exactly the same late-time expansion history, while of course at the same time preserving $\theta_s$ and $z_{\text{eq}}$. Recall in fact that the background expansion history of our IDE model is \textit{exactly} identical to that of a $w$CDM model, i.e.\ a model where the DE component has a constant EoS $w_x \neq -1$, provided the DE EoS is set to $w_x^{\text{EM}}=w_{x,\text{eff}}=w_x+\xi/3$, and the present-day DM and DE densities are readjusted according to the mapping discussed earlier in Eqs.~(\ref{eq:expansionmatchingc},\ref{eq:expansionmatchingx}), which implies the following map:
\begin{align}
\omega_c^{\text{EM}}&=\omega_c^{\text{SM}}+\frac{\xi}{3w_x+\xi}\omega_x^{\text{SM}}\,,
\label{eq:emomegac}\\
\omega_x^{\text{EM}}&=\frac{3w_x}{3w_x+\xi}\omega_x^{\text{SM}}\,.
\label{eq:emomegax}
\end{align}
where we have suppressed the obvious $_0$ subscript, as is common practice when dealing with physical density parameters. In practice, we need only shift $\omega_c$ according to Eq.~(\ref{eq:emomegac}), since the Friedmann equation closure relation, together with the requirement of spatial flatness, then consistently fixes $\omega_{x,0}$. We then compare the IDE$_{\text{SM}}^A$ model against the $w$CDM$_{\text{EM}}^A$ model, for which $w_x^{\text{EM}} \simeq -1.1$, and similarly the IDE$_{\text{SM}}^B$ model against the $w$CDM$_{\text{EM}}^B$ model, for which $w_x^{\text{EM}} \simeq -0.9$. With this mapping (see Tab.~\ref{tab:referencecosmologies}), the two pairs of cosmologies are guaranteed to share the same values of $H_0$, $\theta_s$, and $z_{\text{eq}}$, as well as the same background expansion history and cosmological distances at \textit{all} redshifts (there is thus, by construction, no expansion-matched counterpart to Fig.~\ref{fig:fpbackground} and Fig.~\ref{fig:smbackground}, as these would be completely trivial). This provides a much more stringent comparison than the scale-matched one. In fact, any residual difference we find in the observables \textit{cannot} be ascribed to a different expansion history or line-of-sight geometry. Instead, these differences will necessarily reflect the different decomposition of the same dark sector background into DM and DE, related to the well-known dark degeneracy~\cite{Kunz:2007rk,vonMarttens:2019ixw,Petri:2025swg}, and how this affects the resulting evolution of their perturbations. In other words, this expansion-matched comparison will isolate the IDE signatures which cannot be reproduced by a non-interacting $w$CDM cosmology via a change in the background expansion alone. Strictly speaking, these signatures are not purely perturbative: besides obeying different perturbation equations, the two models contain the same total dark sector energy density $\rho_c+\rho_x$, while $\rho_c$ and $\rho_x$ differ individually, so the amount of clustering DM at a given redshift is different. Any surviving difference will reflect both this effect and the different perturbation equations, both of which are inseparable consequences of the DM-DE energy transfer: in this sense, we identify the differences which survive this expansion-matched comparison as \textit{genuine signatures of IDE}.

We start by looking at the CMB temperature power spectrum, shown in the upper sub-panels of Fig.~\ref{fig:ema} (for model class $A$) and Fig.~\ref{fig:emb} (for model class $B$), respectively the expansion-matched counterparts of Fig.~\ref{fig:sma} and Fig.~\ref{fig:smb}. For both model classes, the CMB power spectrum is basically indistinguishable from its $w$CDM counterpart on intermediate and small scales. This agreement follows from the same arguments presented in the scale-matched case, which we therefore do not repeat here. We have explicitly verified that the gravitational driving and acoustic oscillation dynamics, as captured by the relevant transfer functions, are virtually identical. As in the scale-matched comparison, the remaining differences between the two models are due to the modified late ISW effect: despite sharing the same $H(z)$, the different clustering properties of the two models result in a different late-time evolution of the gravitational potentials, and hence a different contribution to the late ISW tail. The differences we find are extremely small and isolated to multipoles $\ell \lesssim 30$, which are completely cosmic variance-dominated, making them of little observational relevance. In summary, we therefore find that, once $\theta_s$, $z_{\text{eq}}$, and the complete expansion history are matched, the CMB temperature power spectra of IDE and its non-interacting $w$CDM counterpart are essentially indistinguishable.

We then consider the CMB lensing potential power spectrum, shown in the lower left sub-panels of Fig.~\ref{fig:ema} and Fig.~\ref{fig:emb}. We recall that, unlike primary CMB anisotropies, CMB lensing is sensitive to the different post-recombination evolution of the two dark sectors, as it probes an appropriately weighted line-of-sight integral of the Weyl potential $(\phi+\psi)/2$. The CMB lensing potential power spectrum is therefore indirectly sensitive to the entire evolution between recombination and the present day. We also recall that the CMB lensing kernel contains various geometrical quantities, such as the comoving distance to the surface of last scattering and to all intermediate redshifts. Since these are identical by construction in the expansion-matched comparison, any residual difference in $C_L^{\phi\phi}$ inevitably originates from differences in the evolution of the Weyl potential along the line of sight. The reason is once more that, even at fixed total dark sector density, the division into DM and DE is different due to the dark degeneracy, and the growth of DM perturbations is altered by the interaction. This changes the evolution of gravitational potentials even in the absence of background/geometrical differences between the two models. For both model classes, the IDE lensing spectrum differs only modestly from its $w$CDM counterpart, with model-dependent, scale-dependent residuals at the percent or sub-percent level. Qualitatively, we can understand this from Eqs.~(\ref{eq:growth},\ref{eq:stauxi}): for $\xi<0$, the reduced effective friction and additional source term enhance the growth of DM perturbations, whereas the Weyl potential is suppressed in IDE because of the smaller late-time DM density. The competition between these two effects produces the scale-dependent residuals observed in Figs.~\ref{fig:ema} and \ref{fig:emb}, although their detailed shapes follow from the complete numerical solution. These scale-dependent residuals can be considered genuine IDE signatures, which cannot be reproduced solely through a modification of the background expansion. However, given the sensitivity of current and near-future CMB lensing measurements, they are well beyond the reach of observations.

\begin{figure*}[htbp]
\centering
\includegraphics[width=0.99\textwidth]{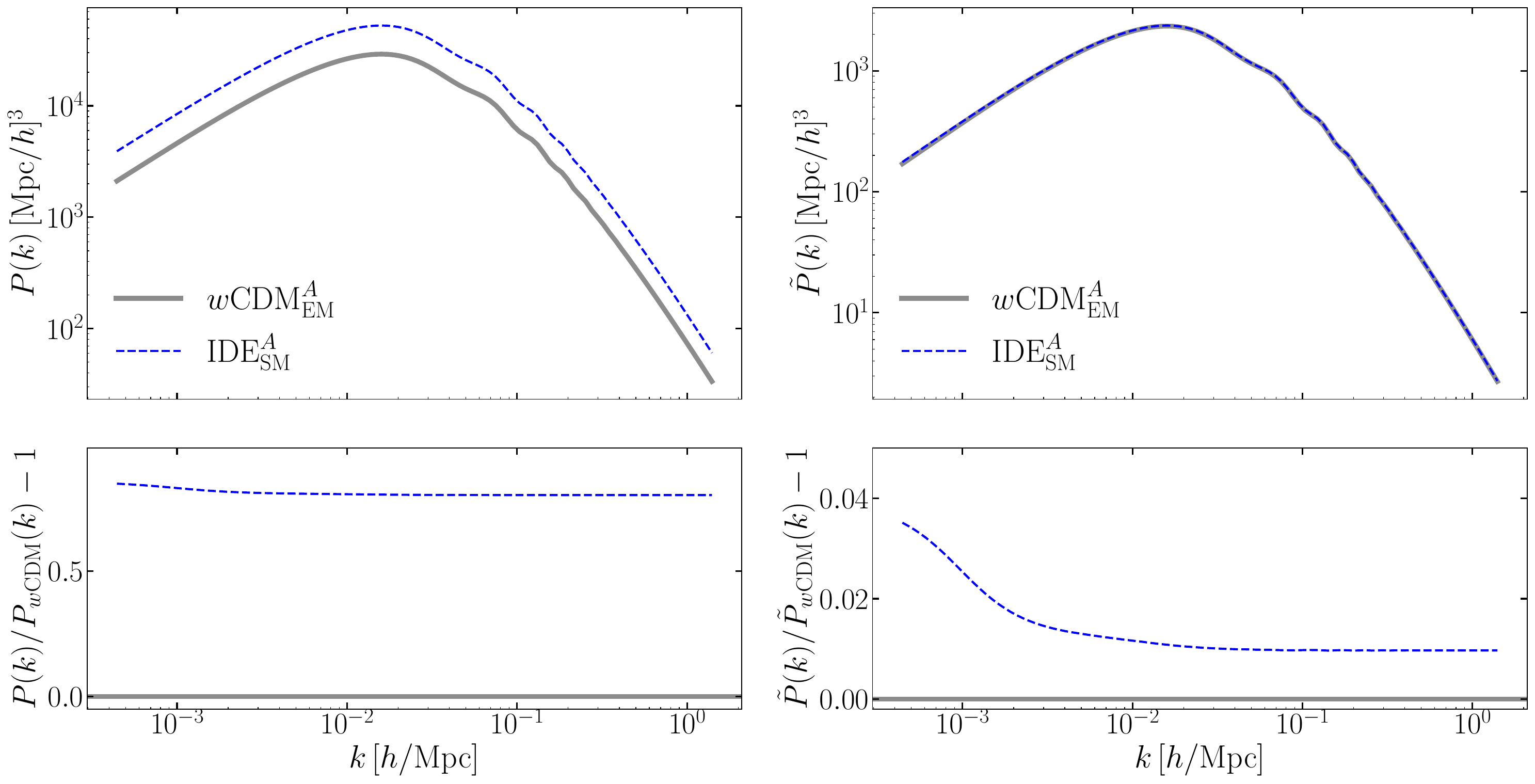}
\caption{Role of the different values of $\Omega_m$ in the expansion-matched matter power spectra of IDE$_{\text{SM}}^A$ and $w$CDM$_{\text{EM}}^A$. In all sub-panels, the gray solid and blue dashed curves correspond to $w$CDM$_{\text{EM}}^A$ and IDE$_{\text{SM}}^A$ respectively. \textit{Left panel}: matter power spectrum $P_m(k)$, with the fractional difference of IDE$_{\text{SM}}^A$ relative to $w$CDM$_{\text{EM}}^A$ shown in the lower sub-panel (this is identical to the lower right panel of Fig.~\ref{fig:ema}). \textit{Right panel}: same as the left column, but for the ``density-weighted matter power spectrum'' $\widetilde P(k) \equiv \Omega_m^2P_m(k)$ and the corresponding fractional difference. Note that the values of $\Omega_m$ are different for the IDE$_{\text{SM}}^A$ and $w$CDM$_{\text{EM}}^A$ models. We see that the rescaling accounts for most of the large amplitude difference observed in the left sub-panel, with only a percent-level, scale-dependent residual remaining.}
\label{fig:pkom}
\end{figure*}

The most significant differences between the expansion-matched IDE and $w$CDM models show up in the matter power spectrum, shown in the lower right sub-panels of Fig.~\ref{fig:ema} and Fig.~\ref{fig:emb}. Since the two models share the same values of $k_{\text{eq}}$ and $H_0$ by construction, the turnover occurs at the same scale even when reported in units of $h\,{\text{Mpc}}^{-1}$. The amplitudes of the two spectra, however, are very different. For instance, for model class $A$, we find that $P_m(k)$ for the IDE model is enhanced by $\approx 75\%$ on large scales, with a large enhancement persisting on smaller scales. While one may be tempted to interpret this as a smoking-gun signature of IDE, we believe that some care is required in doing so. We recall that the matter power spectrum is conventionally defined in terms of the fractional matter density perturbation $\delta_m=\delta \rho_m/\rho_m$, whose normalization explicitly depends on the mean matter density of the cosmology in question. However, as we can see from Tab.~\ref{tab:referencecosmologies}, the expansion-matched IDE and $w$CDM models have different values of $\Omega_m$, which is smaller for IDE as a consequence of the energy transfer from DM to DE. Therefore, even for comparable values of $\delta\rho_m$, the two models can have very different matter power spectra. One may wish to separate the effect of this different normalization from the effects due to the altered growth of density perturbations sourcing the gravitational potential. For this purpose, it is useful to consider the quantity $\widetilde P(k) \equiv \Omega_m^2P_m(k)$, which we call a ``density-weighted matter power spectrum''. We recall that, at $z=0$ (as we are considering here) and on sub-horizon scales, the Poisson equation tells us that $\Omega_m\delta_m$ is indeed the matter contribution to the gravitational potential source (recall that $H_0$ is identical for both models). This comparison is shown explicitly for model class $A$ in Fig.~\ref{fig:pkom}.

As we can see from the right column of Fig.~\ref{fig:pkom}, the $\Omega_m^2$ weighting accounts for the overwhelming majority of the large difference between the unweighted matter power spectra in the left column. The fractional difference in $\widetilde P(k)$ is a few percent on the largest scales shown, decreases to $\approx 1\%$ for $k\gtrsim10^{-2}\,h\,\text{Mpc}^{-1}$, and remains approximately at this level on smaller scales. The residual is therefore positive but modest and mildly scale-dependent. This can be understood qualitatively from Eqs.~(\ref{eq:growth},\ref{eq:stauxi}): for $\xi<0$, the reduced effective friction and additional source term enhance the growth of DM perturbations, partially compensating for the smaller DM density. As in the case of the CMB lensing power spectrum, the detailed scale dependence of this effect follows from the complete numerical solution. We note that CMB lensing probes the line-of-sight projection of the Weyl potential. On the sub-horizon scales which are most relevant to the CMB lensing signal, where we can neglect DE perturbations and anisotropic stress, the matter contribution to the Weyl potential is proportional to $\rho_m\delta_m$, rather than $\delta_m$ alone. Although Fig.~\ref{fig:pkom} shows the corresponding combination at $z=0$, whereas CMB lensing integrates it along the line of sight, this explains why the large difference in $P_m(k)$ does not translate into a comparably large CMB lensing signal. A legitimate question at this point is: can we regard the differences in $P_m(k)$ due to the different values of $\Omega_m$ as a genuine IDE signature? In our opinion, the answer is yes and no. Yes, since the different values of $\Omega_m$ are a direct consequence of the interaction within our expansion-matched prescription: insofar as one accepts this matching, different values of $\Omega_m$ are unavoidable. No, because the order-unity enhancement of $P_m(k)$ is largely due to the different value of $\Omega_m$ used to normalize $\delta_m$, rather than to a large modification of the matter contribution to the gravitational potential source. The density-weighted spectrum $\widetilde P(k)$, which we stress should be interpreted merely as a diagnostic, factors out this normalization and provides a clearer view of the differences in the matter contribution to the gravitational potential source. As Fig.~\ref{fig:pkom} shows, these differences remain at the percent level.

Finally, comparing Fig.~\ref{fig:ema} and Fig.~\ref{fig:emb}, we see that the picture discussed above is essentially unchanged when switching between model classes, whose expansion-matched expansion histories lie on either side of $w_x^{\text{EM}}=-1$. More specifically, we observe that the CMB temperature power spectra are virtually identical, whereas the CMB lensing residuals show the same small scale-dependent differences discussed previously. The matter power spectra show the same strong enhancement, which we find is again consistent with the different values of $\Omega_m$. While the precise amplitude of these differences does depend somewhat on $w_x$, the overall ``hierarchy'' of signatures (i.e.\ in order of smallest to largest signatures, CMB temperature power spectrum, CMB lensing power spectrum, matter power spectrum) does not, demonstrating that our main results are not due to our having chosen an IDE model very close to an interacting vacuum model.

\section{Discussion}
\label{sec:discussion}

We now critically discuss the implications of our results. The main take-away lesson from our analysis is that the (apparent) size and physical interpretation of IDE's cosmological signatures depend \textit{crucially} on the quantities which are being held fixed during the comparison against the reference model, be this $\Lambda$CDM or the non-interacting $w$CDM model. Simply switching on the interaction while keeping $H_0$ and the other $\Lambda$CDM parameters fixed, as in the \textit{fixed-parameter} comparison, and as has been done in earlier comparison plots in the literature, leads to very large changes across all the main observables considered, especially the CMB temperature power spectrum. While these changes should not be interpreted as genuine IDE signatures, the fixed-parameter comparison is nevertheless useful for understanding the raw response of cosmological observables to the interaction, and identifying which interaction-induced changes need to be compensated by shifts in the other cosmological parameters in order to maintain a good fit to the data. In this sense, it provides a useful starting point for understanding the parameter degeneracies at play. However, if our aim is to identify the observationally relevant and ultimately genuine signatures of IDE given the constraints imposed by the data, these compensations need to be taken into account. Once $\theta_s$ and $z_{\text{eq}}$ are preserved, as in our \textit{scale-matched} comparison, the CMB temperature power spectrum is essentially restored (up to small changes in the late ISW effect), as we have explicitly shown at the level of transfer functions. At the same time, significant differences remain in the background expansion history and clustering of matter, together with smaller changes in the CMB lensing power spectrum. When we match the complete expansion history to that of a non-interacting $w$CDM model, as in our \textit{expansion-matched} analysis, we remove all remaining signatures which can be reproduced by the background evolution alone. Aside from the small changes related to the late ISW effect, the surviving signatures are a percent-level scale-dependent change in the CMB lensing power spectrum and a large enhancement of the matter power spectrum. As shown in Fig.~\ref{fig:pkom}, most of the latter is accounted for by the different value of $\Omega_m$, itself an inevitable consequence of IDE due to the energy exchange between DM and DE. Once we factor out this normalization effect by considering the density-weighted spectrum $\widetilde P(k) \equiv \Omega_m^2P_m(k)$, only a percent-level, mildly scale-dependent residual remains. We have shown that these signatures, and in particular their ``hierarchy'' across the three comparisons, are qualitatively unchanged between model classes $A$ and $B$: this shows that our key conclusions are not specific to the vacuum-like limit of model class $A$.

One important lesson we wish to stress is that the background signatures identified in our scale-matched analysis (see Fig.~\ref{fig:smbackground}) should \textit{not} be dismissed as non-genuine simply because they disappear (by construction) in our expansion-matched analysis. They are indeed genuine consequences of the DM-DE interaction, as the interaction modifies the evolution of the DM and DE densities, and therefore the background expansion rate and cosmological distances, precisely in the way shown in Fig.~\ref{fig:smbackground}. In this sense, these are exactly the signatures of IDE which are being probed by DESI BAO and other late-time background measurements, and which underlie recent claims for possible evidence of IDE~\cite{Giare:2024smz,Li:2024qso,Silva:2025hxw,Pan:2025qwy,Li:2026xaz}. At the same time, as we have demonstrated in Sec.~\ref{sec:genuineeffects}, the background expansion rate of our model is identical to that of a non-interacting $w$CDM model, after appropriate remapping of the physical DM and DE densities and the DE EoS. Therefore, while background cosmological data can favor this IDE model over $\Lambda$CDM, they \textit{cannot} by themselves tell whether the deviation from $\Lambda$CDM originates from an interacting DE component or from an uncoupled dynamical DE component, because of the dark degeneracy. The fact that these background signatures can be exactly mimicked \textit{does not make them non-genuine, only non-unique}.

Another important lesson we learned from this analysis is that, for the IDE model considered here, the primary CMB acts essentially as a (high-precision) calibrator of the relevant physical scales ($\theta_s$ and $z_{\text{eq}}$), rather than as a direct probe of the interaction, despite the latter entering the Einstein-Boltzmann equations. In fact, once $\theta_s$ and $z_{\text{eq}}$ are preserved, the relevant transfer functions at the time of recombination are virtually indistinguishable from their $\Lambda$CDM counterparts (see the upper sub-panels of Fig.~\ref{fig:smanatomycmb}). The small remaining differences are limited to the late ISW tail (see the lower sub-panels of Fig.~\ref{fig:smanatomycmb}), and are therefore of little observational relevance, given the large cosmic variance uncertainty. On the other hand, the CMB lensing power spectrum provides complementary information. Since it probes the integrated post-recombination evolution of the Weyl potential, signatures of the interaction survive even after expansion matching. The remaining scale-dependent percent-level signatures are relatively small (see the lower left sub-panels of Fig.~\ref{fig:ema} and Fig.~\ref{fig:emb}), and therefore limit the discriminating power of CMB lensing on its own. However, these could be a useful target for multi-probe analyses combining CMB lensing with geometrical and growth of structure measurements. We defer a full investigation of this possibility to future work.

Perhaps somewhat counterintuitively, the above considerations suggest that most of the CMB constraining power for this class of IDE models could be captured by an appropriate compressed CMB likelihood, but not by blindly applying the usual $\Lambda$CDM distance priors. For instance, let us consider the widely used 3-dimensional compressed likelihood on $\{\theta_s^{-1},\omega_b,\omega_c\}$. If one wishes to use a similar parameter basis, we expect that $\omega_c$ should be replaced by $\omega_c^{\text{eff}}=\omega_c+\xi\omega_x/(3w_x+\xi)$, as per Eq.~(\ref{eq:emomegac}), with $\omega_x$ in turn determined by the Friedmann closure relation. Moreover, the appropriate parameter central values and correlations with other compressed quantities would of course need to be consistently recomputed. Similar considerations would hold for related compression schemes making use of the shift parameters $R$ and $l_a$, since $R$ depends on the matter density. Of course, these modified compression schemes would need to be recalibrated against the full CMB likelihood. While we expect the above to be a reasonable compression strategy for the information content of the primary CMB, checking whether this is indeed the case is well beyond the scope of this work, and is left to follow-up work.

Among all the observables we considered, the matter power spectrum shows the largest differences even after expansion matching (see the lower right sub-panels of Fig.~\ref{fig:ema} and Fig.~\ref{fig:emb}), although interpreting these differences requires some care, since the IDE and expansion-matched $w$CDM models have different values of $\Omega_m$. As explicitly demonstrated for model class $A$ in Fig.~\ref{fig:pkom}, most of the large amplitude difference is accounted for by this difference in $\Omega_m$. On sub-horizon scales, the ``density-weighted spectrum'' $\widetilde P(k)\equiv\Omega_m^2P_m(k)$ is proportional to the power spectrum of $\Omega_m\delta_m$, i.e.\ the matter contribution to the source of the gravitational potential. Comparing $\widetilde P(k)$ rather than $P_m(k)$ therefore allows us to factor out the normalization effect associated with the different value of $\Omega_m$, and doing so reveals a positive, mildly scale-dependent residual at the percent level.

That being said, we stress once more that the different value of $\Omega_m$ is by itself a \textit{genuine} prediction of IDE, and arises from the different partition of the total dark sector energy density between DM and DE. However, the enhancement due to the different value of $\Omega_m$ cannot on its own be interpreted as a modification of the growth dynamics. All these considerations make a determination of $\Omega_m$ which is as high-fidelity and model-independent as possible a key ingredient in testing IDE.~\footnote{Interesting possibilities in this sense are measurements of the gas mass fraction in relaxed massive galaxy clusters~\cite{Allen:2002sr}, or the use of internal properties of individual galaxies~\cite{Villaescusa-Navarro:2022twv}.} Besides IDE, such a determination would be especially timely also in view of the emerging ``CMB-BAO tension'', which within $\Lambda$CDM manifests itself as a tension in $\Omega_m$~\cite{Baryakhtar:2024rky,Pedrotti:2024kpn,Colgain:2024mtg,Weiner:2026sfm,Shlivko:2026jxa,Schoneberg:2026buf,Colgain:2026ryf}.

Let us briefly discuss the possibility of observing the signatures we identified in the matter power spectrum. On the largest scales, this corresponds to a quasi-scale-independent rescaling of $P_m(k)$ (see the lower right sub-panels of Fig.~\ref{fig:ema} and Fig.~\ref{fig:emb}): the signal is therefore expected to be largely degenerate with the linear galaxy bias $b_1$, since a shift in this parameter can almost completely absorb the signal of interest. This is true at the level of the monopole. However, this degeneracy can be broken by utilizing other probes beyond the monopole, such as the quadrupole, whose different dependence on the linear bias and the growth rate through redshift-space distortions can help disentangle the two and isolate signatures of IDE. Another possibility to break this degeneracy is to carry out a multi-probe analysis using probes which have a different sensitivity to the linear bias, e.g.\ cross-correlations between the CMB lensing convergence field and the galaxy samples in question (see e.g.\ Refs.~\cite{Giusarma:2018jei,Tanseri:2022zfe}), or cross-correlations with weak lensing. At any rate, in order to reliably calibrate the linear galaxy bias and extract the large-scale IDE signal in galaxy clustering, multi-probe analyses which go beyond the galaxy power spectrum monopole alone are clearly necessary. Additional constraining power may come from much smaller scales ($k \gtrsim 0.3\,h\,{\text{Mpc}}^{-1}$). However, this is precisely the regime where our linear Einstein-Boltzmann treatment gradually ceases to be valid, and effects such as non-linear evolution and baryonic feedback become important. We have refrained from na\"{i}vely applying prescriptions calibrated on $\Lambda$CDM or $w$CDM (e.g.\ \texttt{Halofit}) to our linear power spectrum, as we cannot assume that these remain valid in the presence of interactions between DM and DE. For this reason, we caution against overinterpreting the signatures shown in Fig.~\ref{fig:ema} and Fig.~\ref{fig:emb} on sufficiently non-linear scales. At the same time, we stress that the difficulty in modeling non-linear clustering in IDE models is technical rather than fundamental: $N$-body simulations of IDE models exist, and recent works have shown that the non-linear evolution can actually be modeled directly~\cite{Silva:2024ift,Zhai:2026uwr}. Future analyses aiming to extract the non-linear signal could make use of these dedicated simulations, eventually combined with emulators, and/or appropriately extended prescriptions based on the Effective Field Theory of LSS (concerning this last point, see however the caveats outlined in Ref.~\cite{Nunes:2022bhn}). A quantitative investigation of these non-linear signatures is of course well beyond the scope of this work, and we therefore defer it to a follow-up study.

Overall, our analysis has helped clarify the physical signatures of IDE. The importance of carrying out a ``controlled dissection'', such as that performed in our analysis, is well appreciated in other areas of cosmology, such as the study of massive neutrinos~\cite{Lesgourgues:2013sjj,Vagnozzi:2019utt}. However, we feel that it has received much less attention in the context of IDE (despite the significant attention received by these models in the literature), where the effects of the DM-DE interaction are often illustrated by na\"{i}vely varying $\xi$, which, as we have shown, in the best case overestimates the signatures of the model (and in the worst case misidentifies them). While parameter estimation analyses implicitly account for all the parameter shifts and compensations of our three-layer analysis, solely focusing on parameter estimation, as is usually done, can obscure the physical origin of the resulting constraints. We stress that we do not intend to detract from the merits of parameter estimation analyses, which remain of paramount importance. Rather, our controlled analyses are intended to complement rather than replace them. The expansion-matched analysis also explicitly highlights the well-known dark degeneracy~\cite{Kunz:2007rk,vonMarttens:2019ixw,Petri:2025swg}, i.e.\ the fact that background observables on their own cannot distinguish interacting and non-interacting models producing the same expansion history. In order to confirm the presence of a dark sector interaction, identifying the signatures we have found in observables primarily sensitive to the evolution of perturbations and gravitational potentials is therefore crucial: this is of paramount importance at the present time, when tentative indications of dark sector interactions have been reported in light of recent background cosmological observations~\cite{Giare:2024smz,Li:2024qso,Silva:2025hxw,Pan:2025qwy,Li:2026xaz}.

While we believe our work is an important step towards understanding the genuine cosmological signatures of IDE, there are a number of limitations which should be kept in mind. Firstly, we have focused on a specific phenomenological choice for the energy exchange rate, i.e.\ $Q=\xi{\cal H}\rho_x$, with constant DE EoS $w_x$ and an energy-momentum transfer four-vector parallel to the DM four-velocity. While this is among the most widely studied phenomenological IDE models, especially in light of cosmological tensions and the possible evidence for dynamical DE, we cannot automatically assume that our results extend to other interaction rates or momentum-transfer prescriptions. Similar considerations hold for our choice of interaction strength $\xi=-0.3$, which we have adopted simply as a reasonable benchmark to magnify the observable effects, while remaining within the range explored by cosmological constraints. Moreover, we remind the reader that our choice of energy exchange rate is phenomenological. It could be very interesting to extend our analysis to IDE models based on concrete field theories, although this would need to be done on a case-by-case basis~\cite{Pan:2020zza,Aboubrahim:2024cyk,Zhang:2025dwu,Li:2026xaz,Abdalla:2026sis}. Finally, as stressed earlier, our analysis is based on the linear Einstein-Boltzmann system, and the results obtained in the mildly non-linear regime should therefore be considered with caution. While genuine signatures of IDE persist on smaller scales, quantifying and exploiting them requires dedicated non-linear modeling.

\section{Conclusions}
\label{sec:conclusions}

Motivated by the renewed observational interest in interacting dark energy (IDE) models, in light not only of persisting cosmological tensions, but especially of the latest DESI results, this work aims to address a simple but surprisingly subtle question: \textit{what are the genuine cosmological signatures of IDE}? The overwhelming majority of studies on IDE have (quite understandably) focused on parameter estimation, but much less attention has been devoted to identifying the true physical origin of the obtained constraints. For instance, comparisons between IDE and $\Lambda$CDM are often done na\"{i}vely, by varying the interaction strength while keeping $H_0$ and other $\Lambda$CDM parameters fixed. We have argued that this obscures IDE's genuine signatures, as it does not distinguish between features intrinsic to the DM-DE interaction and features which are absorbed by shifts in other cosmological parameters in order to keep a number of relevant physical scales fixed. Focusing on the simple, phenomenological, and widely studied IDE model with energy exchange rate $Q=\xi{\cal H}\rho_x$ (with $\rho_x$ the DE density), we address the key question of this work by carrying out a sequence of controlled comparisons. We begin with the usual na\"{i}ve fixed-parameter comparison for illustrative purposes, before moving on to a scale-matched comparison, where shifts in the other cosmological parameters preserve the acoustic angular scale $\theta_s$ and the redshift of matter-radiation equality $z_{\text{eq}}$. In the third, expansion-matched comparison, we compare IDE against a non-interacting $w$CDM model with exactly the same expansion history, but a different partition between the DM and DE components. We implement this procedure for a number of cosmological observables, including the background geometry (expansion rate and distances), the CMB temperature power spectrum (carefully studying its physical sources and the relevant transfer functions), the CMB lensing power spectrum, and the matter power spectrum. We represent this sequence of controlled comparisons through the metaphor of peeling an onion (see Fig.~\ref{fig:onion}), with each step removing one layer and the core representing signatures of IDE which cannot be attributed either to shifts in accurately measured physical scales or to differences in the background expansion history: it is these surviving signatures which we identify as distinctive signatures of the DM-DE interaction.

We find that the large effects observed in the na\"{i}ve fixed-parameter comparisons provide a misleading picture of IDE's real observational signatures, as the large shifts in $\theta_s$ and $z_{\text{eq}}$ lead to large, observationally excluded shifts in the CMB temperature acoustic peaks. Once we preserve $\theta_s$ and $z_{\text{eq}}$ in our scale-matched analysis (through shifts in $H_0$ and $\omega_c$, and indirectly $\Omega_m$), the CMB temperature power spectrum is essentially completely restored, with differences well below the percent level on most scales. We have explicitly shown that this occurs because the transfer functions for the Weyl potential, effective photon monopole, and baryon velocity are essentially identical to their $\Lambda$CDM counterparts, especially around recombination, leading to near-identical CMB temperature power spectrum sources. The CMB temperature power spectrum therefore acts as a high-precision calibrator of the relevant physical scales, with IDE-induced modifications of the perturbations playing essentially no role in the primary CMB temperature anisotropies once these scales are matched. Small changes appear in the late ISW tail at low multipoles, due to the different post-recombination evolution of the Weyl potential; however, these changes are too small and appear in a region which is too strongly dominated by cosmic variance to be observationally relevant. These findings remain true in our expansion-matched analysis, where we compare the IDE model against a non-interacting $w$CDM model with the same expansion history, but a different value of $\omega_c$ (and therefore of $\Omega_m$). The CMB lensing power spectrum shows a modest but characteristic scale-dependent signature of IDE, with the power spectrum being suppressed at the percent level at low multipoles and enhanced at a similar level at high multipoles. We instead observe much larger differences in the matter power spectrum. However, we have shown that the difference can be almost entirely accounted for by the different values of $\Omega_m$ which IDE inevitably requires: considering the density-weighted spectrum $\widetilde P(k) \equiv \Omega_m^2P_m(k)$, we find only a modest, percent-level, mildly scale-dependent residual. We stress that the different value of $\Omega_m$ required is itself a genuine consequence of the energy exchange between DM and DE within our expansion-matched comparison; however, the large enhancement observed in $P_m(k)$ as a result of the different values of $\Omega_m$ cannot on its own be interpreted as a modification of the growth dynamics.

Looking at the broader implications of our results, the background signatures surviving our scale-matched analysis are genuine consequences of IDE. These are precisely the signatures which are being targeted, and have potentially been detected, by background probes such as DESI BAO. However, they cannot by themselves distinguish between a DM-DE interaction and a non-interacting DE model with appropriately remapped physical densities, a manifestation of the well-known dark degeneracy. For this reason, in order to conclusively establish a preference for dark sector interactions over non-interacting explanations, it is essential to look for correlated signatures in the clustering of LSS tracers. In this sense, we view growth-geometry consistency tests as a potentially particularly promising way forward. As we have argued, the success of such a program requires a number of ingredients, ranging from a high-fidelity determination of $\Omega_m$ (whose impact and importance, at the time of writing, would extend well beyond IDE), which is needed to separate its dominant contribution to the amplitude shift in $P_m(k)$ from the remaining percent-level scale-dependent residual, to multi-probe analyses to help break the degeneracies with galaxy bias. Our results also suggest that an appropriate compressed CMB likelihood can be constructed for IDE models, provided $\omega_c$ and its correlations are appropriately remapped, although a dedicated study would be required to confirm this point. More generally, we point out that the controlled comparison strategy we have carried out here is obviously not restricted to our specific interaction model, nor to IDE models alone. This point is especially timely, as we are witnessing a growing number of tentative indications for late-time departures from $\Lambda$CDM, with a wide range of extensions (especially related to the dark sector) being confronted with increasingly precise cosmological data: as new cosmological data that may confirm these deviations arrives~\cite{SimonsObservatory:2018koc,SimonsObservatory:2019qwx,Euclid:2024yrr}, the same strategy of systematically removing layers can be applied to these extensions in order to uncover their genuine physical signatures. Finally, as possible indications for DM-DE interactions become more frequent, we stress that these claims should be tested against appropriately matched non-interacting scenarios: the signatures surviving this comparison are those which would distinguish dark sector interactions from modifications to the background expansion history, and are precisely the smoking-gun signatures one should seek in other probes, such as LSS clustering. We leave a detailed study of these points to future follow-up work.

\begin{acknowledgments}
\noindent M.A.S.\ and S.V.\ acknowledge support from the Istituto Nazionale di Fisica Nucleare (INFN) through the Commissione Scientifica Nazionale 4 (CSN4) Iniziativa Specifica ``Quantum Fields in Gravity, Cosmology and Black Holes'' (FLAG). E.D.V.\ is supported by a Royal Society Dorothy Hodgkin Research Fellowship. This publication is based upon work from the COST Action CA21136 ``Addressing observational tensions in cosmology with systematics and fundamental physics'' (CosmoVerse), supported by COST (European Cooperation in Science and Technology).
\end{acknowledgments}

\bibliographystyle{apsrev4-1}
\bibliography{anatomyide}

\end{document}